\documentclass[aps,prd,reprint,groupedaddress,10pt,nofootinbib]{revtex4-1}

\usepackage{hyperref}  
\usepackage{amsmath,amssymb}
\usepackage{mathtools}
\usepackage{fancyvrb}

\allowdisplaybreaks[1]
\usepackage{cprotect}
\newcommand{\dr}{\mathrm{d}}
\usepackage{color}

\newcommand{\intn}{\mathrm{int}}
\newcommand{\tr}{\mathrm{tr}}

\newcommand{\diag}{\mathrm{diag}}

\renewcommand{\vec}[1]{\boldsymbol{#1}}

\newcommand{\Slash}[1]{\ooalign{\hfil/\hfil\crcr$#1$}}

\def\simge{\mathrel{
   \rlap{\raise 0.511ex \hbox{$>$}}{\lower 0.511ex \hbox{$\sim$}}}}
\def\simle{\mathrel{
   \rlap{\raise 0.511ex \hbox{$<$}}{\lower 0.511ex \hbox{$\sim$}}}}
\def\bigs{\mathrel{
   \rlap{\raise 0.531ex \hbox{$>$}}{\lower 0.531ex \hbox{$<$}}}}

\usepackage{booktabs}
\usepackage[normalem]{ulem}  
\usepackage{multirow}

\begin{document}
\preprint{}

\title{Open charm and bottom meson-nucleon potentials {\`a} la 
the nuclear force}
\author{Yasuhiro~Yamaguchi}
\email[]{yamaguchi@hken.phys.nagoya-u.ac.jp}
\affiliation{Department of Physics, Nagoya University, Nagoya 464-8602, Japan}
\affiliation{Advanced Science Research Center, Japan Atomic Energy Agency (JAEA), Tokai 319-1195, Japan}
\author{Shigehiro~Yasui}
\email[]{yasuis@keio.jp}
\affiliation{Research and Education Center for Natural Sciences, Keio University, Hiyoshi 4-1-1, Yokohama, Kanagawa 223-8521, Japan}
\author{Atsushi~Hosaka}
\email[]{hosaka@rcnp.osaka-u.ac.jp }
\affiliation{Research Center for Nuclear Physics (RCNP), Ibaraki, Osaka 567-0047, Japan}
\affiliation{Advanced Science Research Center, Japan Atomic Energy Agency, Tokai, Ibaraki 319-1195, Japan}
\affiliation{Theoretical Research Division, Nishina Center, RIKEN, Hirosawa, Wako, Saitama 351-0198, Japan}

\date{\today}

\begin{abstract}
We discuss the interaction of 
an open heavy meson ($\bar{D}$ and $\bar{D}^{\ast}$ for charm or $B$ and $B^{\ast}$ for bottom) and a nucleon ($N$)
by considering the $\pi$, $\sigma$, $\rho$, and $\omega$ exchange potentials.
We construct a potential model by respecting chiral symmetry for light quarks and spin symmetry for heavy quarks. 
Model parameters are adjusted by referring the phenomenological nuclear (CD-Bonn) potentials reproducing the low-energy $NN$ scatterings.
We show that the resulting interaction may accommodate
$\bar{D}N$ and $BN$ bound states with 
quantum numbers  $I(J^{P})=0(1/2^{-})$, and $1(1/2^{-})$.
We find that, in the present potential model, 
the $\pi$ exchange potential plays an important role for  
the isosinglet channel, while the $\sigma$ exchange potential does for 
the isotriplet one.
\end{abstract}

\maketitle

\section{Introduction} \label{sec:introduction}

Studies of exotic hadrons, such as $X$, $Y$, $Z$, $P_{c}$, $X_{cc}$, $T_{cc}$, and so on, have revealed novel properties of multiquark systems with
heavy flavors of charm and bottom~\cite{Swanson:2006st,Voloshin:2007dx,Brambilla:2010cs,Brambilla:2014jmp,Chen:2016qju,Hosaka:2016pey,Lebed:2016hpi,Esposito:2016noz,Ali:2017jda,Olsen:2017bmm,Guo:2017jvc,Liu:2019zoy,Brambilla:2019esw,Yamaguchi:2019vea,Chen:2022asf}. 
One of the most important problems in exotic hadrons is the inter-hadron interactions.
In the present paper, we focus on the interaction between a nucleon $N$ and an open-heavy meson, a $\bar{D}$ ($\bar{D}^{\ast}$) meson or a $B$ ($B^{\ast}$) meson, which is intimately relevant to the formation of pentaquarks.
Such 
an
interaction is also relevant for
heavy-flavored exotic nuclei as bound states formed by a multiple number of baryons~\cite{Hosaka:2016ypm}.
Recently the ALICE collaboration in LHCb has reported the first experimental 
study
of the $\bar{D}N$ interaction which was measured through the correlation functions from proton-proton collisions~\cite{alicecollaboration2022study}.
Further development of studying the interaction between a nucleon $N$ and an open-heavy meson should be awaited.

One of the efficient theoretical analyses
can be performed
systematically with the basis on the heavy-quark effective theory. 
This is an effective theory of QCD, where a charm (bottom) quark is approximately regarded as a particle with an infinitely heavy mass 
$m_{Q}\rightarrow\infty$. 
In this limit, there appears the heavy-quark spin (HQS) symmetry, i.e., the SU(2) spin symmetry, 
as 
in
the nonrelativistic limit.
This symmetry stems from the decoupling of the heavy quark from 
light degrees of freedom with the suppressed magnetic interaction, i.e., 
the spin-flip interaction.
The HQS symmetry puts conditions on the spin structure of interaction vertices not only in the quark-gluon dynamics but also in the hadron dynamics.

The HQS symmetry is seen in the observed approximate degeneracy in masses of $\bar{D}$ and $\bar{D}^{\ast}$ ($B$ and $B^{\ast}$) mesons.
Also,
the HQS symmetry constrains the structure of
the inter-hadron interaction in the channel-coupled $\bar{D}N$ and $\bar{D}^{\ast}N$ ($BN$ and $B^{\ast}N$) systems.
For example, it was shown that the approximate degeneracy in $\bar{D}$ and $\bar{D}^{\ast}$ mesons 
increases 
the
attractive interaction strength between a nucleon and a $\bar{D}$ meson through the box diagram $\bar{D}N \rightarrow \bar{D}^{\ast}N \rightarrow \bar{D}N$ in the 
second-order
perturbative process~\cite{Haidenbauer:2007jq}.
This mechanism is different from the conventional approach based on the SU(4) flavor symmetry~\cite{Lutz:2005vx,Hofmann:2005sw} and the quark-meson coupling model~\cite{Haidenbauer:2007jq,Fontoura:2012mz,Carames:2012bd,Carames:2017pfl}.
The role of the HQS symmetry is shown to be important  
by including all the coupled channels of $\bar{D}N$ and $\bar{D}^{\ast}N$ ($BN$ and $B^{\ast}N$).
Hereafter we will introduce the short notations $P$ and $P^{\ast}$ 
corresponding to 
$\bar{D}$ and $\bar{D}^{\ast}$ ($B$ and $B^{\ast}$), respectively.
We employ
$P^{(\ast)}$ to denote either 
$P$ or $P^{\ast}$.
In such a framework, we consider the coupled channels of $PN$ and $P^{\ast}N$ 
and study
the interaction between a $P^{(\ast)}$ meson and a nucleon, denoted by $PN$-$P^{\ast}N$.

In the literature, the $PN$-$P^{\ast}N$ interactions were 
introduced
by the one-pion exchange potential (OPEP) with the constraint conditions induced by the HQS symmetry~\cite{Cohen:2005bx,Yasui:2009bz,Yamaguchi:2011xb,Yamaguchi:2011qw,Yasui:2013vca,Yamaguchi:2014era}.
The analysis of the $PN$-$P^{\ast}N$ systems showed the possible existence of composite states: bound states 
below
the $\bar{D}N$ ($BN$) threshold~\cite{Cohen:2005bx,Yasui:2009bz}, and Feshbach resonant states in the 
continuum
region slightly 
below
the $\bar{D}^{\ast}N$ ($B^{\ast}N$) threshold~\cite{Yamaguchi:2011xb,Yamaguchi:2011qw}.
In the heavy quark limit, these two states are regarded as the doublet states in the mixed bases by $PN$ and $P^{\ast}N$ in terms of the HQS symmetry. 
Such HQS multiplets have been studied for negative parity~\cite{Yasui:2009bz,Yamaguchi:2011xb} and 
positive parity~\cite{Yamaguchi:2011qw}
states.

In HQS symmetry, it is important to realize that the $PN$-$P^{\ast}N$ interactions can be 
provided 
by the interaction
between the nucleon and the light quark $q$ in $P^{(\ast)}$.
Thus, the $PN$-$P^{\ast}N$ interactions can be regarded effectively as the ``$qN$" interaction.
This would be a generalization of the conventional nuclear force ($NN$) to the force between a {\it light quark} and a nucleon with different baryon numbers.
Such 
an
idea enables us to 
construct the $PN$-$P^{\ast}N$ potential
from the $qN$ potential, with reference to the $NN$ potential in detail.  
It was shown 
that the $PN$-$P^{\ast}N$ interaction can be expressed by 
the $qN$ interaction by applying the unitary transformations~\cite{Yasui:2013vca,Yamaguchi:2014era}.

In the present work, we reconstruct the $PN$-$P^{\ast}N$ potential, where we refer to the phenomenological nuclear potential, the CD-Bonn potential~\cite{Machleidt:2000ge}. 
In the framework of the CD-Bonn potential, the nuclear force is described by the $\pi$, $\rho$, 
$\omega$,
and $\sigma$
exchanges.
It is known that the $\sigma$ exchange is important to reproduce the phase shifts in $NN$ scatterings for isospin singlet and triplet channels simultaneously.
In fact, the $\pi$, $\rho$, and $\omega$-exchange potentials are not enough for the fitting to the observed data of $NN$ scatterings.
In reference to the CD-Bonn potential,
thus 
we also introduce the middle-range force by the $\sigma$ exchange potential
in addition to the $\pi$, $\rho$, and $\omega$ potentials in the $PN$-$P^{\ast}N$ interaction which were discussed by the previous studies~\cite{Yasui:2009bz,Yamaguchi:2011xb,Yamaguchi:2011qw}.
As introduced in the CD-Bonn potential, the parameters of the $\sigma$ exchange have 
different values 
between the isosinglet and isotriplet channels.
Considering $\bar{D}N$-$\bar{D}^{\ast}N$ and $BN$-$B^{\ast}N$ systems
with the reconstructed potentials,
we discuss the possible existence of 
bound states,
as 
discussed in~\cite{Yasui:2009bz,Yamaguchi:2011xb,Yamaguchi:2011qw}.

The paper is organized as the followings.
In Sec.~\ref{sec:potential}, we introduce the potentials for $PN$ and $P^{\ast}N$ in terms of the $\pi$, $\sigma$, $\rho$, and $\omega$ exchanges.
We give an analysis for the $\sigma$ exchange potential which is newly introduced in the present study.
We present in details the calculation process of the derivation of the potential, because we include some corrections for the potential forms derived in our previous works.
In Sec.~\ref{sec:numerical_results}, we present the numerical results for the scattering lengths in the $PN$ and $P^{\ast}N$ potentials and the binding energies for the bound states.
The final section is devoted to our conclusion and prospects for future studies.

\section{Formalism} \label{sec:formalism}

\subsection{Construction of $PN$ and $P^{\ast}N$ potentials} \label{sec:potential}

\subsubsection{OPEP}

Let us consider the $PN$-$P^{\ast}N$ states of 
$J^{P}=1/2^{-}$
with a total angular momentum $J$ and parity $P$.
$PN$ and $P^{\ast}N$ components in those states are represented by
\begin{align}
 PN(^2{S}_{1/2}), P^{\ast}N (^2{S}_{1/2}), P^{\ast}N(^4{D}_{1/2}) .
 \label{eq:particle_basis_1/2^-}
\end{align}
Here the notation $^{2S+1}L_{J}$ in the parentheses stands for the combination of the total spin $S$ and 
 the relative angular momentum $L$ for a given $J$.
In view of the HQS 
symmetry, 
the wave functions given above are decomposed into the product of a heavy antiquark $\bar{Q}$ and a light component 
``$l$". 
Here ``$l$" is nonperturbatively composed of the light quarks ($q$) and gluons ($g$) inside the $PN$-$P^{\ast}N$ state. 
Such 
a light component may be schematically denoted by $qqqq$, because it should be a composite state of the light quark $q$ in $P$ or $P^{\ast}$ and 
the three quarks $qqq$ in the nucleon $N$. 
This is the special case of the so-called 
brown muck
which was introduced in the early days when the 
heavy quark effective theory (HQET)
was constructed.\footnote{In the present setting, the brown muck is regarded to have the special component $qN$ in $qqqq$.}

The idea of the light composite state
leads to the mass degeneracy of the $PN$-$P^{\ast}N$ states with different $J^P$, such as $J^{P}=1/2^{-}$ and $3/2^-$
by taking the heavy quark limit, 
because the spin-dependent interaction between the heavy antiquark ($\bar{Q}$) and the brown muck ($qqqq$) is suppressed by $1/m_{Q}$ with the heavy quark mass $m_{Q}$.
The mass degeneracy of the $PN$-$P^{\ast}N$ states have been studied in Refs.~\cite{Yasui:2013vca,Yamaguchi:2014era,Hosaka:2016ypm}.

For the interaction in the $PN$-$P^{\ast}N$ systems, we adopt the meson-exchange potential between $P^{(\ast)}$ and $N$.
We consider the one-pion exchange potential (OPEP)  as the long-range force.
We also consider 
the $\sigma$-meson exchange potentials and 
the $\rho$ and $\omega$-meson exchange potentials as the middle-range force.

Let us first explain the derivation  of the 
OPEP in details as an illustration.
In constructing the OPEP, we need the information of the interaction vertices of $\pi$ and $P^{(\ast)}$ and those of $\pi$ and $N$.
For the $\pi PP^{\ast}$ and $\pi P^{\ast}P^{\ast}$ vertices, we employ the heavy meson effective theory (HMET) satisfying the HQS as well as chiral symmetry~\cite{Manohar:2000dt,Casalbuoni:1996pg}.
Notice the absence of the $\pi PP$ vertex due to the parity conservation. 

For heavy mesons $P$ and $P^{\ast}$, we define the effective field $H_{a}$ being a superposition of a heavy pseudoscalar meson and a vector meson as
\begin{align}
   H_{\alpha} = \bigl( P_{\alpha}^{\ast \mu}\gamma_{\mu}+P_{\alpha}\gamma_5 \bigr)\frac{1-\Slash{v}}{2},
\label{eq:H_field_def}
\end{align}
where the subscripts $\alpha=\pm1/2$ represent the isospin components (up and down) in the light quark components.
$P_{\alpha}$ and $P_{\alpha}^{\ast \mu}$ denote the pseudoscalar and vector meson fields, respectively.
The relative phase of $P_{\alpha}^{\ast \mu}$ and $P_{\alpha}$ is arbitrary, and the present choice is adopted for the convenience in representing the $PN$-$P^{\ast}N$ potential as it will be shown later.
Here $v^{\mu}$ ($\mu=0,1,2,3$) is the 
four velocity 
of the heavy meson (heavy antiquark) satisfying $v_{\mu}v^{\mu}=1$ and $v^{0}>0$.
We notice that $(1-\Slash{v})/2$ is the operator for projecting out the positive-energy component in the heavy antiquark $\bar{Q}$ and discarding the negative-energy component.
The complex conjugate of $H_{\alpha}$ is defined by $\bar{H}_{\alpha} = \gamma_{0} H_{\alpha}^{\dagger} \gamma_{0}$.
The effective field $H_{\alpha}$ transforms as 
$H_{\alpha} \rightarrow U_{\alpha\beta} H_{\beta}S^\dagger$
under the heavy-quark spin and chiral symmetries.
Here $S \in \mathrm{SU}(2)_{\mathrm{spin}}$ represents the transformation operator for the heavy-quark spin and $U_{\alpha\beta}=U_{\alpha\beta}(L,R)$ is a function in the nonlinear representation of chiral symmetry with $L \in \mathrm{SU}(2)_{\mathrm{L}}$ and $R \in \mathrm{SU}(2)_{\mathrm{R}}$ for light up and down flavors.

In terms of $H_{\alpha}$ defined by Eq.~\eqref{eq:H_field_def}, the interaction Lagrangian for the $\pi P^{(\ast)}P^{(\ast)}$ vertex is given by
\begin{align}
{\cal L}_{\pi HH} =   
 i  g_{\pi}  
\mbox{tr} \bigl( H_{\alpha}\bar{H}_{\beta}\gamma_{\mu}\gamma_{5} A^{\mu}_{\beta\alpha} \bigr),
\label{eq:LpiHH}
\end{align}
where the axial current $A^{\mu}_{\beta\alpha}$ by pions is defined by
$A^{\mu} = \bigl( \xi^{\dag} \partial^{\mu}\xi - \xi \partial^{\mu} \xi^{\dag} \bigr)/2$ with the nonlinear representation
\begin{align}
      \xi = \exp\biggl(i\frac{\vec{\tau}\!\cdot\!\vec{\pi}}{2f_{\pi}}\biggr),
\end{align}
with the pion decay constant $f_{\pi}=94$ MeV.
The pion field is defined by $\vec{\pi}=(\pi_{1},\pi_{2},\pi_{3})$ with $\pi^{\pm}=(\pi_{1}\mp i\pi_{2})/\sqrt{2}$ for charged pions and 
$\pi^{0}=\pi_{3}$
for a neutral pion.
Notice that the matrix $A^{\mu}$ 
is transformed by $A^{\mu} \rightarrow U A^{\mu} U^{\dag}$ in the nonlinear representation of chiral symmetry.
Thus we confirm that the interaction Lagrangian~\eqref{eq:LpiHH} is invariant under both the 
HQS and chiral symmetries.
The coupling constant $g_\pi=0.59$
in Eq.~\eqref{eq:LpiHH} is 
determined from the decay width of $D^{\ast -} \rightarrow D^{-}\pi^{0}$ observed by experiments~\cite{ParticleDataGroup:2020ssz}.
We note that $g_\pi$ is nothing but the quark axial coupling  $g_A^q$ whose value 
looks smaller than what is naively expected, $g_A^q = 1$~\cite{Weinberg:1990xm}.  
The small value is understood by considering corrections
due to quark's relativistic motion inside hadrons as discussed in detail for baryon decays~\cite{Arifi:2022ntc}.
There are uncertainties for choosing the signs of the coupling constants in $\bar{D}$ ($\bar{D}^\ast$) and $B$ ($B^\ast$). 
In the present study, we 
assume 
that the $\pi$, $\sigma$, $\rho$, and $\omega$ mesons couple to the light constituent quarks in the heavy mesons as well as in the nucleons. 
In this scheme, we can consider that the signs of these meson couplings for the light quarks in the heavy meson are the same as for the nucleon, 
because both have the same light 
(up and down) 
constituent quarks according to the conventional quark model.

Below we consider the 
frame in which the heavy meson is 
at rest
and set $v^{\mu} = (1, \vec{0})$ in Eq.~\eqref{eq:LpiHH}. Thus we obtain the $\pi P^{(\ast)}P^{(\ast)}$ vertices:
\begin{align}
   {\cal L}_{\pi P^{\ast}P^{\ast}}
&=
 \frac{ig_{\pi}}{f_{\pi}} 
   \varepsilon_{\nu\rho\mu\sigma}v^{\nu}
   P_{\beta}^{\ast\rho\dag} \bigl(\vec{\tau} \!\cdot\! \partial^{\mu}\vec{\pi}\bigr)_{\beta\alpha} P_{\alpha}^{\ast\sigma},
\label{eq:piHH_vertex_1} \\ 
   {\cal L}_{\pi P^{\ast}P}
&=
 i \frac{ig_{\pi}}{f_{\pi}} 
   P_{\beta\mu}^{\ast\dag} \bigl(\vec{\tau} \!\cdot\! \partial^{\mu}\vec{\pi}\bigr)_{\beta\alpha} P_{\alpha},
\label{eq:piHH_vertex_2} \\ 
   {\cal L}_{\pi PP^{\ast}}
&=
 i \frac{ig_{\pi}}{f_{\pi}} 
   P_{\beta}^{\dag} \bigl(\vec{\tau} \!\cdot\! \partial^{\mu}\vec{\pi}\bigr)_{\beta\alpha} P_{\alpha\mu}^{\ast}.
\label{eq:piHH_vertex_3}
\end{align}

We introduce the interaction Lagrangian of a pion and a nucleon in the axial-vector coupling
\begin{align}
   {\cal L}_{\pi NN}
& =
    \frac{g_{A}^{N}}{2f_{\pi}} 
   \bar{\psi}\gamma_{\mu}\gamma_{5} \vec{\tau}\!\cdot\!\partial^{\mu}\vec{\pi}\psi . 
\label{eq:LpiNN_av}
\end{align}
Here $\psi=(\psi_{+1/2},\psi_{-1/2})^{T}$ 
with the isospin components 
$\psi_{+1/2}$ and $\psi_{-1/2}$ for a proton and a neutron, respectively.
The value of 
$g_{A}^{N}$
is given by 
the Goldberger-Treiman relation
\begin{align}
 \frac{g_{A}^{N}}{f_\pi} = \frac{g_{\pi NN}}{m_N} ,
\end{align}
and $g_{\pi NN}^{2}/4\pi=13.6$ from the phenomenological nuclear potential in Ref.~\cite{Machleidt:2000ge} (see also Ref.~\cite{Machleidt:1987hj}).
We adopt the values of the coupling constants and the cutoff parameters by referring the parameters in the CD-Bonn potential.
The nuclear potentials used in the present study are explained in Appendix~\ref{sec:NN_potential}.

With the interaction vertices \eqref{eq:LpiHH} and \eqref{eq:LpiNN_av}, we construct the OPEP between $P^{(\ast)}$ and $N$ \cite{Yasui:2009bz,Yamaguchi:2011xb,Yamaguchi:2011qw}.
We show the demonstration to derive the potential for the simple model in Appendix~\ref{sec:deriving_potential}.
The OPEP includes three channels: $P^{\ast}N \rightarrow P^{\ast}N$, $P^{\ast}N \rightarrow PN$, and $PN \rightarrow P^{\ast}N$.
We notice that the $PN \rightarrow PN$ process is absent as a direct process due to the prohibition of the $\pi PP$ vertex, and that the $PN$-$PN$ interaction is indirectly supplied by multi-step process stemming from the mixing of $PN$ and $P^{\ast}N$~\cite{Yasui:2009bz,Yamaguchi:2011xb,Yamaguchi:2011qw}.
The OPEPs for $P^{\ast}N$-$P^{\ast}N$, $P^{\ast}N$-$PN$, and $PN$-$P^{\ast}N$ are given by
\begin{widetext}
\begin{align}
   V_{\pi}^{P^{\ast}N\text{-}P^{\ast}N}(\vec{r})
&=
   { G_{\pi} }
    \Bigl(
        T(r;m_{\pi})
         \bigl(
               3(\vec{T}\!\cdot\!\hat{\vec{r}}) 
                 (\vec{\sigma}\!\cdot\!\hat{\vec{r}}) 
             - \vec{T}
                \!\cdot\!\vec{\sigma} 
                \bigr)
       + C(r;m_{\pi})
         \vec{T} 
         \!\cdot\!\vec{\sigma} 
   \Bigr)
   \vec{\tau}^{H} 
   \!\cdot\!
   \vec{\tau}^{N}, 
\label{eq:barDN_V_PS_potentials_3A} \\[0.5em] 
   V_{\pi}^{P^{\ast}N\text{-}PN}(\vec{r})
&=
  - {G_{\pi}} 
    \Bigl(
        T(r;m_{\pi})
         \bigl(
               3(\vec{\epsilon}^{\ast}  
                  \!\cdot\!\hat{\vec{r}}) (\vec{\sigma}\!\cdot\!\hat{\vec{r}})  
             - \vec{\epsilon}^{\ast}  
                \!\cdot\!\vec{\sigma} 
          \bigr)
      + C(r;m_{\pi})
         \vec{\epsilon}^{\ast} 
         \!\cdot\!\vec{\sigma} 
   \Bigr)
   \vec{\tau}^{H}  
     \!\cdot\!
   \vec{\tau}^{N},  
\label{eq:barDN_V_PS_potentials_3B} \\[0.5em] 
   V_{\pi}^{PN\text{-}P^{\ast}N}(\vec{r})
&=
  -  {G_{\pi} }  
   \Bigl(
        T(r;m_{\pi})
         \bigl(
               3(\vec{\epsilon}  
                  \!\cdot\!\hat{\vec{r}}) (\vec{\sigma}\!\cdot\!\hat{\vec{r}}) 
             - \vec{\epsilon} 
                  \!\cdot\!\vec{\sigma}  
          \bigr)
      + C(r;m_{\pi})
         \vec{\epsilon} 
          \!\cdot\!\vec{\sigma} 
   \Bigr)
   \vec{\tau}^{H}  
     \!\cdot\!
   \vec{\tau}^{N},  
\label{eq:barDN_V_PS_potentials_3C}
\end{align}
\end{widetext}
with the coefficient
\begin{align}
   {G_{\pi} } =
 \frac{1}{3}{\frac{1}{2}}\frac{g_{\pi NN}}{2m_{N}}\frac{g_{\pi}}{f_{\pi}}.
\end{align}
We notice that the coefficient $1/2$ is necessary due to the normalization factor of the wave functions,
which was missing in Refs.~\cite{Yasui:2013vca,Yamaguchi:2014era,Hosaka:2016ypm}.
The derivation of the OPEP is shown 
in Appendix~\ref{sec:derivation_OPEP} in details.
The functions $C(r;m)$ and $T(r;m)$ are defined by
\begin{align}
   C(r;m)
&=
   \frac{m^{2}}{4\pi}
   \frac{1}{r}
   \nonumber \\ & \times 
   \biggl(
         e^{-mr}
      + \frac{\Lambda_{H}^{2}-m^{2}}{\Lambda_{N}^{2}-\Lambda_{H}^{2}}
         e^{-\Lambda_{N}r}
      + \frac{\Lambda_{N}^{2}-m^{2}}{\Lambda_{H}^{2}-\Lambda_{N}^{2}}
         e^{-\Lambda_{H}r}
   \biggr),
\label{eq:C_def}
\\ 
   T(r;m)
&=
 \frac{1}{4\pi} 
   \Biggl(
            m^{2}
            \biggl(
                  \frac{1}{r}
               + \frac{3}{mr^{2}}
               + \frac{3}{m^{2}r^{3}}
            \biggr)
            e^{-mr}
            \nonumber \\ & \hspace{1em} 
         + \Lambda_{N}^{2}
            \biggl(
                  \frac{1}{r}
               + \frac{3}{\Lambda_{N}r^{2}}
               + \frac{3}{\Lambda_{N}^{2}r^{3}}
            \biggr)
            \frac{\Lambda_{H}^{2}-m^{2}}{\Lambda_{N}^{2}-\Lambda_{H}^{2}}
            e^{-\Lambda_{N}r}
            \nonumber \\ & \hspace{1em} 
         + \Lambda_{H}^{2}
            \biggl(
                  \frac{1}{r}
               + \frac{3}{\Lambda_{H}r^{2}}
               + \frac{3}{\Lambda_{H}^{2}r^{3}}
            \biggr)
            \frac{\Lambda_{N}^{2}-m^{2}}{\Lambda_{H}^{2}-\Lambda_{N}^{2}}
            e^{-\Lambda_{H}r}
   \Biggr),
\label{eq:T_def}
\end{align}
with $m=m_{\pi}$, respectively, as functions of an 
interdistance 
$r=|\vec{r}|$ for $\vec{r}$ being the relative coordinate vector between $P^{(\ast)}$ and $N$.
The detailed information to derive the potentials are presented in Appendix~\ref{sec:derivation_OPEP}.
Notice that the values of the cutoff parameters $\Lambda_{H}$ 
($H=\bar{D}, B$)
and $\Lambda_{N}$ are dependent on the species of the exchanged 
light meson,  
e.g., the $\pi$ meson.
Originally, 
$C(r,m)$ and $V(r,m)$
are defined by
\begin{align}
 &C(r;m)
   =\int\frac{d^3 \vec{q}}{(2\pi)^3}
 \frac{m^2}{\vec{q}^{\,\,2}+m^2}e^{i\vec{q}\cdot\vec{r}}F(\vec{q};m)\,
 , \label{Cpote}\\
 &S_{\cal O}(\hat{\vec{r}})
 T(r;m)
 =\int\frac{d^3 \vec{q}}{(2\pi)^3}
 \frac{-\vec{q}^{\,\,2}}{\vec{q}^{\,\,2}+m^2}S_{\cal O}(\hat{\vec{q}})e^{i\vec{q}\cdot\vec{r}}F(\vec{q};m)
 , \label{Tpote}
\end{align}
 for the central and tensor parts, respectively,
with $\hat{\vec{q}}=\vec{q}/|\vec{q}|$. 
We note that the contact term in the central part is neglected.
The dipole-type form factor is given by
\begin{align}
   F(\vec{q};m)
= \frac{\Lambda_{H}^2-m^2}{\Lambda_{H}^2+|\vec{q}|^2}
   \frac{\Lambda_{N}^2-m^2}{\Lambda_{N}^2+|\vec{q}|^2},
\label{eq:dipole_form_factor}
\end{align}
which is normalized at $q^2=m^2$ with a four-momentum $q$.
The cutoff parameters $\Lambda_{H}$ and $\Lambda_N$ would correspond to the inverse of the spatial sizes of hadrons.
See the derivations in Appendix~\ref{sec:derivation_OPEP} for more details.
In Eqs.~\eqref{eq:barDN_V_PS_potentials_3B} and \eqref{eq:barDN_V_PS_potentials_3C}, we define the polarization vectors
 $\vec{\epsilon}^{\,(\lambda)}$ ($\vec{\epsilon}^{\,(\lambda)*}$) for the incoming (outgoing) $P^\ast$ meson with the polarization $\lambda=0,\, \pm1$.
The explicit forms of $\vec{\epsilon}^{\,(\lambda)}$ can be represented by
\begin{align}
   \vec{\epsilon}^{\,(\pm)} = \frac{1}{\sqrt{2}}\bigl(\mp1,-i,0\bigr), \quad
   \vec{\epsilon}^{\,(0)} = (0,0,1),
\label{eq:polarization_vector}
\end{align}
by choosing the positive direction in the $z$ axis for the helicity $\lambda=0$.
As 
for the spin-one operator for the $P^{\ast}$ meson in Eq.~\eqref{eq:barDN_V_PS_potentials_3A}, we define $\vec{T}=(T_{1},T_{2},T_{3})$ by $(T_{i})_{\lambda'\lambda} \equiv - i\varepsilon_{ijk} \epsilon_{j}^{(\lambda')\ast} \epsilon_{k}^{(\lambda)}$ ($i,j,k=1,2,3$):
\begin{align}
   T_{1}
&=
   \frac{1}{\sqrt{2}}
   \left(
   \begin{array}{ccc}
    0 & 1 & 0 \\
    1 & 0 & 1 \\
    0 & 1 & 0  
   \end{array}
   \right),
\quad
   T_{2}
=
   \frac{1}{\sqrt{2}}
   \left(
   \begin{array}{ccc}
    0 & -i & 0 \\
    i & 0 & -i \\
    0 & i & 0  
   \end{array}
   \right),
\nonumber \\ 
   T_{3}
&=
   \left(
   \begin{array}{ccc}
    1 & 0 & 0 \\
    0 & 0 & 0 \\
    0 & 0 & -1  
   \end{array}
   \right),
\label{eq:T_matrices}
\end{align}
satisfying the commutation relation $[T_{i},T_{j}]=i\varepsilon_{ijk}T_{k}$ as the generators of the spin symmetry.
We define the tensor operators $S_{\vec{\epsilon}}(\hat{\vec{r}})$ and $S_{\vec{T}}(\hat{\vec{r}})$ by $S_{\vec{\cal O}}(\hat{\vec{r}})=3(\vec{\cal O}\!\cdot\!\hat{\vec{r}})(\vec{\sigma}\!\cdot\!\hat{\vec{r}})-\vec{\cal O}\!\cdot\!\vec{\sigma}$ with $\hat{\vec{r}}=\vec{r}/r$ for $\vec{{\cal O}}=\vec{\epsilon}$ and $\vec{T}$.
Here $\vec{\sigma}$ are the Pauli matrices acting on the nucleon spin, and 
$\vec{\tau}^{H}_{\beta_{1}\alpha_{1}}$ and $\vec{\tau}^{N}_{\beta_{2}\alpha_{2}}$
with $\alpha_{i},\beta_{i}=\pm1/2$ are the isospin Pauli operators for $P^{(\ast)}$ ($i=1$) and $N$ ($i=2$), respectively.

Using the basis of the $J^{P}=1/2^{-}$ channel in Eq.~\eqref{eq:particle_basis_1/2^-}, 
we represent the OPEPs \eqref{eq:barDN_V_PS_potentials_3A}, \eqref{eq:barDN_V_PS_potentials_3B}, and \eqref{eq:barDN_V_PS_potentials_3C} by the matrix forms,
\begin{align}
V_{1/2^-}^{\pi} &=
\left(
\begin{array}{ccc}
 0 & \sqrt{3} \, {C}_{\pi} & -\sqrt{6} \, {T}_{\pi}  \\
\sqrt{3} \, {C}_{\pi} & -2 \, {C}_{\pi} & -\sqrt{2} \, {T}_{\pi} \\
-\sqrt{6} \, {T}_{\pi} & -\sqrt{2} \, {T}_{\pi} & {C}_{\pi} - 2\, {T}_{\pi}
\end{array}
\right), 
\label{eq:Hamiltonian_pi_1/2-} 
\end{align}
where we define 
${C}_{\pi}=G_{\pi} C(r;m_{\pi})$ and ${T}_{\pi} = G_{\pi} T(r;m_{\pi})$ 
for short notations.
In Eq.~\eqref{eq:Hamiltonian_pi_1/2-}, 
we confirm that the mixing between $PN$ and $P^{\ast}N$ are represented by the off-diagonal parts including the tensor potentials.
These tensor potentials induce the strong mixing by different angular momenta, leading to the strong attractions at short-range scales.
Thus, the mixing of $PN$ and $P^{\ast}N$ is important to switch on the strong attraction.
This is analogous to the OPEP in the nucleon-nucleon interaction.

\subsubsection{$\sigma$ exchange potential} \label{sec:sigma_exchange}

The interaction Lagrangian for a $\sigma$ meson and a $P^{(\ast)}$ meson 
is given by
\begin{align}
   {\cal L}_{\sigma_{I} HH} = \, &
 -  g_{\sigma_{I}} \tr \bigl( \bar{H} \sigma_{I} H \bigr), 
\end{align}
which leads to the $\sigma P^{(\ast)} P^{(\ast)}$ vertices, 
\begin{align}
{\cal L}_{\sigma_{I} PP} = \, &  2 g_{\sigma_{I}} \left( P^{\dagger} \sigma_{I} P\right) , \\
{\cal L}_{\sigma_{I} P^\ast P^\ast } = \, &  -  2 g_{\sigma_{I}} \left( P^{\ast\mu\dagger} \sigma_{I} P^{\ast}_\mu \right) .
\end{align}
Here we introduce 
the channel-dependent $\sigma_{I}$ meson 
for isospin-singlet $(I=0)$ and isospin-triplet $(I=1)$ channels for the $PN$-$P^{\ast}N$ scatterings, as introduced in the CD-Bonn potential~\cite{Machleidt:2000ge}. 
The parameter of the $\sigma$ exchange potential in the CD-Bonn potential~\cite{Machleidt:2000ge} has the different value for each partial waves, i.e., isospin channels.
Thus,
$\sigma_{I}$ in the present work also has an 
channel-dependent mass 
($m_{\sigma_I}$),  
coupling constant ($g_{\sigma_{I}}$), and cutoff parameter ($\Lambda_{\sigma_{I}}$). 
Using the $\sigma NN$ vertices given by
\begin{align}
   {\cal L}_{\sigma_{I} NN} &= 
 g_{\sigma_{I}NN} \bar{\psi}\sigma_{I}\psi,
\end{align}
we find that the $\sigma$ potentials for $PN$ and $P^{\ast}N$ are obtained by
\begin{align}
   V_{\sigma_{I}}^{PN\text{-}PN}(r)
&=
 - \frac{g_{\sigma_{I}NN}g_{\sigma_{I}}}{m_{\sigma_{I}}^{2}}
   C(r;m_{\sigma_{I}}),
\\ 
   V_{\sigma_{I}}^{P^{\ast}N\text{-}P^{\ast}N}(r)
&=
 - \frac{g_{\sigma_{I}NN}g_{\sigma_{I}}}{m_{\sigma_{I}}^{2}}
   C(r;m_{\sigma_{I}}) ,
\end{align}
where we employ the values of $m_{\sigma_{I}}$ and $g_{\sigma_{I}NN}$ in the CD-Bonn potential, see Appendix~\ref{sec:NN_potential}.
Concerning the values of $g_{\sigma_{I}}$, we choose 
$g_{\sigma_{I}} = g_{\sigma_{I}NN}/3$ by assuming that the coupling of a $\sigma$ meson and a hadron $h=P^{(\ast)}$, $N$ is proportional to the number of the light quarks in the hadron $h$: one 
light quark
in $P^{(\ast)}$ and three 
light quarks
in $N$.
The $\sigma$-exchange potentials are expressed explicitly by
\begin{align}
   V_{1/2^{-}}^{\sigma_{I}}
&=
   \left(
   \begin{array}{ccc}
    C_{\sigma_{I}} & 0 & 0 \\
    0 & C_{\sigma_{I}} & 0 \\
    0 & 0 & C_{\sigma_{I}} 
   \end{array}
   \right),
\end{align}
for the 
basis
by Eq.~\eqref{eq:particle_basis_1/2^-},
where we define the function
\begin{align}
   C_{\sigma_{I}}
=- \frac{g_{\sigma_{I}NN}g_{\sigma_{I}}}{m_{\sigma_{I}}^{2}} C(r;m_{\sigma_{I}}) 
\end{align}
for short notations.

\subsubsection{$\rho$ and $\omega$ exchanges potential}

Finally, we consider the exchange of the vector mesons, $\rho$ and $\omega$, at shorter range.
The  $\rho$ and $\omega$ potentials can be constructed from the $vP^{(\ast)}P^{(\ast)}$ vertices for light vector meson $v$ ($v = \rho$, $\omega$).
Following the previous papers~\cite{Yasui:2009bz,Yamaguchi:2011xb,Yamaguchi:2011qw},
we consider the interaction Lagrangian
\begin{align}
   {\cal L}_{vHH}
= \, &
 i \beta \tr \bigl( \bar{H}_\beta v^{\mu} (\rho_{\mu})_{\beta\alpha} H_\alpha \bigr)
   \nonumber \\ & 
 +  i\lambda \tr \bigl( \bar{H}_\beta \sigma^{\mu \nu} (F_{\mu \nu}(\rho))_{\beta\alpha} H_\alpha \bigr),
\label{eq:vHH_vertex}
\end{align}
by respecting the HQS
symmetry.
The vector meson field is defined by $\rho_{\mu} = i g_V \hat{\rho}_{\mu}/\sqrt{2}$ with $\hat{\rho}_{\mu}$,
\begin{align}
 \hat{\rho}_{\mu} = 
\begin{pmatrix}
 \frac{\rho^0}{\sqrt{2}} + \frac{\omega}{\sqrt{2}} & \rho^+ \\
 \rho^- & - \frac{\rho^0}{\sqrt{2}} + \frac{\omega}{\sqrt{2}}
\end{pmatrix}_{\mu} ,
\end{align}
and $g_V \simeq 5.8 $ the universal vector-meson coupling.
In Eq.~\eqref{eq:vHH_vertex}, the tensor field
 is given by
$F_{\mu \nu}(\rho) = \partial_{\mu} \rho_{\nu} - \partial_{\nu} \rho_{\mu} +[\rho_{\mu}, \rho_{\nu}] $.
The coupling constants are given by $\beta = 0.9$ and $\lambda = 0.56$ 
GeV$^{-1}$
by following Refs.~\cite{Casalbuoni:1996pg,Isola:2003fh}. 
In Ref.~\cite{Isola:2003fh}, $\beta$ was determined by the vector-meson dominance, and $\lambda$ was evaluated by the long distance charming penguin diagrams in the $B$ meson decay process. 
The $vP^{(\ast)}P^{(\ast)}$ vertices are obtained by the Lagrangians~\eqref{eq:vHH_vertex} as
\begin{align}
 {\cal L}_{v P^\ast P^\ast} = \, & 
 -  \beta g_V v_\mu P^{\ast \dagger}_{\beta \nu} (\vec{\tau}\cdot \vec{\rho}\,^\mu)_{\beta\alpha} P^{\ast\nu}_{\alpha}
 \notag\\
 & 
  +  2i\lambda g_V \left(
 P^{\ast\nu\dagger}_{\beta} (\vec{\tau}\cdot \partial_\mu \vec{\rho}_\nu)_{\beta\alpha} P^{\ast\mu}_\alpha
 \right.
 \notag\\
 & \left.  \quad 
 -  P^{\ast\mu\dagger}_{\beta} (\vec{\tau}\cdot \partial_\mu \vec{\rho}_\nu)_{\beta\alpha} P^{\ast\nu}_\alpha
 \right) ,
 \\
 {\cal L}_{v P^\ast P} = \, & 
 2 \lambda g_V \epsilon_{\sigma\rho\mu\nu}v^{\sigma} 
 P^{\ast\rho\dagger}_{\beta}  (\vec{\tau}\cdot \partial^\mu \vec{\rho}^\nu)_{\beta\alpha} P_{\alpha} ,
 \\
 {\cal L}_{v P P^\ast} = \, & 
 2 \lambda g_V \epsilon_{\sigma\rho\mu\nu}v^{\sigma} 
  P_{\beta}^{\dagger} (\vec{\tau}\cdot \partial^\mu \vec{\rho}^\nu)_{\beta\alpha} P^{\ast\rho}_{\alpha} ,
 \\
 {\cal L}_{v P P} = \, & 
 \beta g_V v_\mu P^\dagger_{\beta} (\vec{\tau}\cdot \vec{\rho}^{\mu})_{\beta\alpha} P_{\alpha} .
\end{align}
For the $vNN$ vertex, we use the interaction Lagrangian
\begin{align}
{\cal L}_{vNN} &=
 g_{\rho NN} 
 \bar{\psi} 
 \gamma_{\mu}
\vec{\tau}
\!\cdot\!\vec{\rho}^{\,\mu} 
 \psi
 + \frac{
 f_{\rho NN} 
 }{2m_N} 
 \bar{\psi}
\sigma_{\mu \nu}
 \vec{\tau}\!\cdot\!\partial^{\mu}\vec{\rho}^{\,\nu} 
 \psi
 \nonumber \\
 &\quad
  +  g_{\omega NN}
 \bar{\psi}
 \gamma_{\mu} \omega^{\mu} 
   \psi
 + \frac{
 f_{\omega NN} 
 }{2m_N} 
\bar{\psi}
 \sigma_{\mu \nu} 
 \partial^{\mu} \omega^{\nu}
  \psi , 
\label{eq:vNN_vertex}
\end{align}
for $\vec{\rho}^{\,\mu}=(\rho^{\mu}_{1},\rho^{\mu}_{2},\rho^{\mu}_{3})$ with $\rho_{\pm}^{\mu}=(\rho_{1}^{\mu} \mp i\rho_{2}^{\mu})/\sqrt{2}$ and $\rho_{0}^{\mu}=\rho_{3}^{\mu}$.
The coupling constants are given by $g_{\rho NN}^2/4\pi = 0.84$, $g_{\omega NN}^2/4\pi = 20.0$, 
$f_{\rho NN}/g_{\rho NN}=6.1$, and $f_{\omega NN}/g_{\omega NN}=0.0$~\cite{Machleidt:2000ge} (see also Ref.~\cite{Machleidt:1987hj}).
We leave a comment that the coupling strengths in Eqs.~\eqref{eq:vHH_vertex} and \eqref{eq:vNN_vertex} reflect the number of constituent quarks inside the hadrons. This can be easily checked by the nonrelativistic quark model. 
We should notice, however, that the tensor parts, $\lambda$ and $f_{vNN}$ ($v=\rho$, $\omega$), could be different by some factors from the naive expectations, which would be understood from the composite structures of the constituent quarks.

From Eqs.~\eqref{eq:vHH_vertex} and \eqref{eq:vNN_vertex}, the one-boson exchange potentials are obtained as
\begin{align}
 V_{1/2^-}^v &=
\begin{pmatrix}
 C_v^{\prime} & 2\sqrt{3}C_v & \sqrt{6}T_v \\
 2\sqrt{3} C_v & C_v^{\prime} -4 C_v & \sqrt{2}T_v \\
 \sqrt{6}T_v & \sqrt{2} T_v & C_v^{\prime} +2C_v +2T_v
\end{pmatrix},  
\end{align}
with $v=\rho$, $\omega$ for the $1/2^-$ state in Eq.~\eqref{eq:particle_basis_1/2^-}.
The functions $C_v^{\prime}$, $C_v$, and $T_v$ are defined by 
\begin{align}
 C_{\rho}^{\prime} &= \frac{g_V g_{\rho NN} \beta}{2  
 m_{\rho}^2}
 C(r;m_{\rho})
 \vec{\tau}^{H} \!\cdot\! \vec{\tau}^{N}, \\
 C_{\rho} &= \frac{g_V 
 (g_{\rho NN}+f_{\rho NN}) \lambda 
 }{ 2 
 m_N} \frac{1}{3}
 T(r;m_{\rho})
 \vec{\tau}^{H} \!\cdot\! \vec{\tau}^{N}, \\ 
 T_{\rho} &= \frac{g_V 
 (g_{\rho NN}+f_{\rho NN}) \lambda 
 }{ 2 
 m_N} \frac{1}{3} 
 T(r;m_{\rho})
 \vec{\tau}^{H} \!\cdot\! \vec{\tau}^{N}, \\
 C_{\omega}^{\prime} &= \frac{g_V g_{\omega NN}\beta}{ 2 
 m_{\omega}^2}
 C(r;m_{\omega}),
 \\
 C_{\omega} &= \frac{g_V 
 (g_{\omega NN}+f_{\omega NN})\lambda 
 }{ 2 
 m_N} \frac{1}{3}
 C(r;m_{\omega}),
 \\
 T_{\omega} &= \frac{g_V 
  (g_{\omega NN}+f_{\omega NN}) \lambda
 }{ 2 
m_N} \frac{1}{3}
 T(r;m_{\omega}),
\end{align}
with $\vec{\tau}^{H}$ and $\vec{\tau}^{N}$ being the abbreviations of $\vec{\tau}^{H}_{\beta_{1}\alpha_{1}}$ and $\vec{\tau}^{N}_{\beta_{2}\alpha_{2}}$ for the isospin Pauli operators acting on $P^{(\ast)}$ and $N$, respectively.

\subsection{Total Hamiltonian}

The total Hamiltonian for the $P^{(\ast)}N$ states is given as a sum of the kinetic term and the $\pi$, $\sigma$, $\rho$, and $\omega$ potentials as
\begin{align}
 H_{IJ^{P}} 
 = K_{J^P} + V^{\pi}_{J^P} + V^{\sigma_{I}}_{J^P}
 + V^{\rho}_{J^P} + V^{\omega}_{J^P}.
\end{align}
Here $K_{J^P}$ is the diagonal matrix for the kinetic terms given by
\begin{align}
   K_{1/2^{-}}
&=
   \diag \bigl( K_{0}, K_{0}^{\ast}, K_{2}^{\ast} \bigr),
\end{align}
where each component is defined by
\begin{align}
K_{L} &= - \frac{1}{2\mu}\left( \frac{\partial^2}{\partial r^2 }+ \frac{2}{r} \frac{\partial}{\partial r} - \frac{L(L+1)}{r^2} \right),
\label{eq:kinetic_term_PS} \\ 
K_{L}^{\ast} &= - \frac{1}{2\mu^\ast}\left( \frac{\partial^2}{\partial r^2 }+ \frac{2}{r} \frac{\partial}{\partial r} - \frac{L(L+1)}{r^2} \right),
\label{eq:kinetic_term_V}
\end{align}
for angular momenta $L=0$ and $L=2$.
The reduced masses $\mu=m_{N}m_{P}/(m_{N}+m_{P})$ and $\mu^{\ast}=m_{N}m_{P^{\ast}}/(m_{N}+m_{P^{\ast}})$ are defined with $m_{P}$ and $m_{P^{\ast}}$ being the masses of $P$ and $P^{\ast}$ mesons, respectively.

Concerning the cutoff parameters in the potentials,
we consider $\Lambda_{H}$ in Eq.~\eqref{eq:dipole_form_factor} to be expressed by $\Lambda_{H}=\kappa_{HN}\Lambda_{N}$ where
$\kappa_{HN}$ is the ratio stemming from inverse 
hadron size.
In Refs.~\cite{Yasui:2009bz,Yamaguchi:2011xb,Yamaguchi:2011qw},
we obtained $\kappa_{\bar{D}N}=1.35$ for the $\bar{D}^{(\ast)}N$ potential and $\kappa_{BN}=1.29$ for the $B^{(\ast)}N$ potential.
The same ratios were adopted for the $\rho$ and $\omega$ exchange potentials, and can be applied also 
to the $\sigma$ exchange potential.
In the present study, however, we regard $\kappa_{HN}$ as a free parameter in order to investigate the dependence of the results on the choice of $\kappa_{HN}$ within a range around
$\kappa_{\bar{D}N}=1.35$ and $\kappa_{BN}=1.29$.
The value of $\Lambda_{N}$ is 
determined by modifying 
the cutoffs in the CD-Bonn potential by 
another scale parameter $\kappa_I$ $(I=0,1)$ 
for each isospin channels.
The scale parameter is 
determined by reproducing 
the scattering lengths
of the $NN$ scatterings for $I=1$ and the binding energy of a deuteron for $I=0$, 
where we employ the simplified nuclear potential neglecting 
the massive 
scalar meson, non-local effects and so on in the CD-Bonn potential, 
see Appendix~\ref{sec:NN_potential} in details.
The obtained cutoffs are summarized in Table~\ref{table:mex_parameter}.

\begin{table*}[tbp]  
\caption{\label{table:mex_parameter} Parameters
 of the meson exchange potentials.
 The meson masses are given as the isospin-averaged values. 
 $g_\pi$, $\beta$, $\lambda$, and $g_{\sigma_I}$ are the coupling constants of heavy mesons (see text in details), while $g_{\alpha NN}$ and $f_{\alpha NN}$ are those of a nucleon taken from the CD-Bonn potential~\cite{Machleidt:2000ge}.
 The cutoffs $\Lambda_{\bar{D}}$ and $\Lambda_B$ are shown as typical values for $\Lambda_{\bar{D}}=1.35 \Lambda_N$ and $\Lambda_B=1.29 \Lambda_N$, where $\Lambda_N$ is the nucleon cutoff which is scaled by  
the parameter $\kappa_I$ ($\kappa_0=0.804$ and $\kappa_1=0.773$) from the CD-Bonn potential (see Appendix~\ref{sec:NN_potential} in details).
 }
\begin{center}
\begin{tabular}{cccccccccccccc}
\hline
   \multirow{2}{*}{Mesons $(\alpha)$} & 
     \multirow{2}{*}{Masses [MeV]} & 
	 \multirow{2}{*}{$g_\pi$} & 
	     \multirow{2}{*}{$\beta$} & 
		 \multirow{2}{*}{$\lambda$ [GeV$^{-1}$]} & 
		     \multirow{2}{*}{$g_{\sigma_I}$} & 
		     \multirow{2}{*}{$\frac{g^{2}_{\alpha NN}}{4\pi}$} & 
			 \multirow{2}{*}{$\frac{f_{\alpha NN}}{g_{\alpha NN}}$}  
			 &
			     \multicolumn{2}{c}{$\Lambda_{\bar{D}}$ [MeV]} &
                                \multicolumn{2}{c}{$\Lambda_{B}$ [MeV]} &
                                   \multicolumn{2}{c}{$\Lambda_{N}$ [MeV]}
		 \\
 &&&&&&&
 &
     $I=0$ & $I=1$  &
	     $I=0$ & $I=1$  &
		     $I=0$ & $I=1$  
		 \\
\hline
   $\pi$ & 
     138.04 & 
	 0.59 &
	     --- &
		 --- &
		 --- &
		     13.6 & --- 			    
	       & 1868 
	       & 1795 
		   & 1785 
		   & 1715 
                      & 1384 
                      & 1330 
		       \\
   $\rho$ & 769.68 & 
	 --- &
	     0.9 &
		 0.56 &
		 --- &
		     0.84 & 6.1 
	& 1359 
        & 1306 
          & 1423 
          & 1367 
    	   & 1054 
           & 1013 
		       \\		 
   $\omega$ & 781.94 & 
	 --- &
	     0.9 &
		 0.56 &
		 --- &
	 20 & 0.0 
	& 1629 
        & 1565 
	   & 1557 
	   & 1496 
	     & 1207 
             & 1159 
		       \\
   $\sigma_{0}$ & 350 & 
		 --- &
		 --- &
		 --- &		 
		      0.849406  &
			 0.51673 & --- 	     	    
	       & 2715 
	       & ---
		   & 2594 
	           & ---
                     & 2011 
 	             & ---
		 \\
   $\sigma_{1}$ & 452 & 
	 --- &
	 --- &
         --- & 		 
		    { 2.35276 } &
		     3.96451 & --- 
			& --- 
			& 2609 
				 & --- 
				 & 2493 
                                     & --- 
				     & 1932 
	     \\
\hline
\end{tabular}
\end{center}
\end{table*}

\section{Numerical results} \label{sec:numerical_results}

\begin{figure*}[t]
\begin{center}
\begin{tabular}{cc}
 (a) $\bar{D}N\, (I=0)$& (b) $\bar{D}N \, (I=1)$ \\
 \includegraphics[scale=0.5]{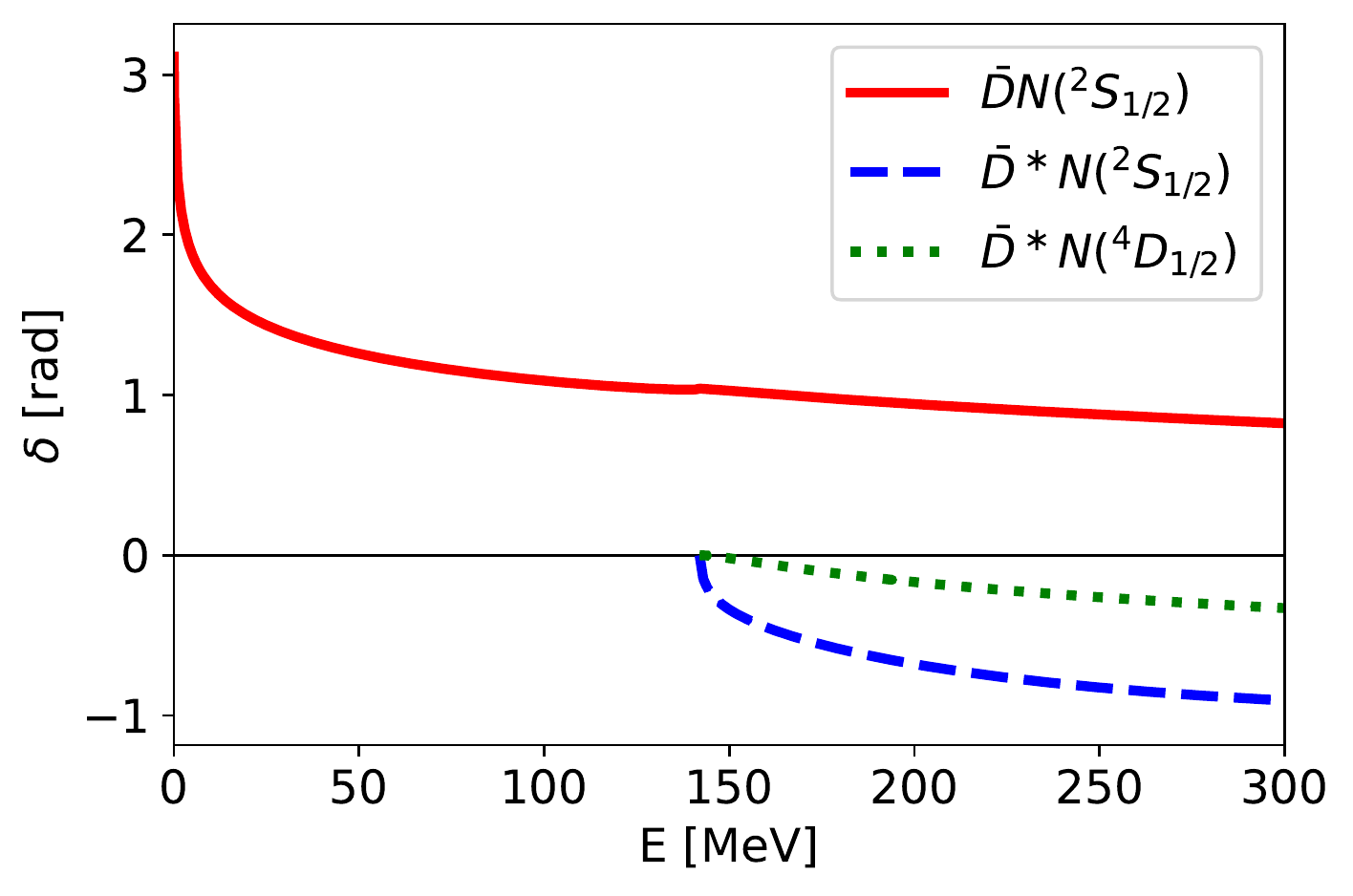} &
\includegraphics[scale=0.5]{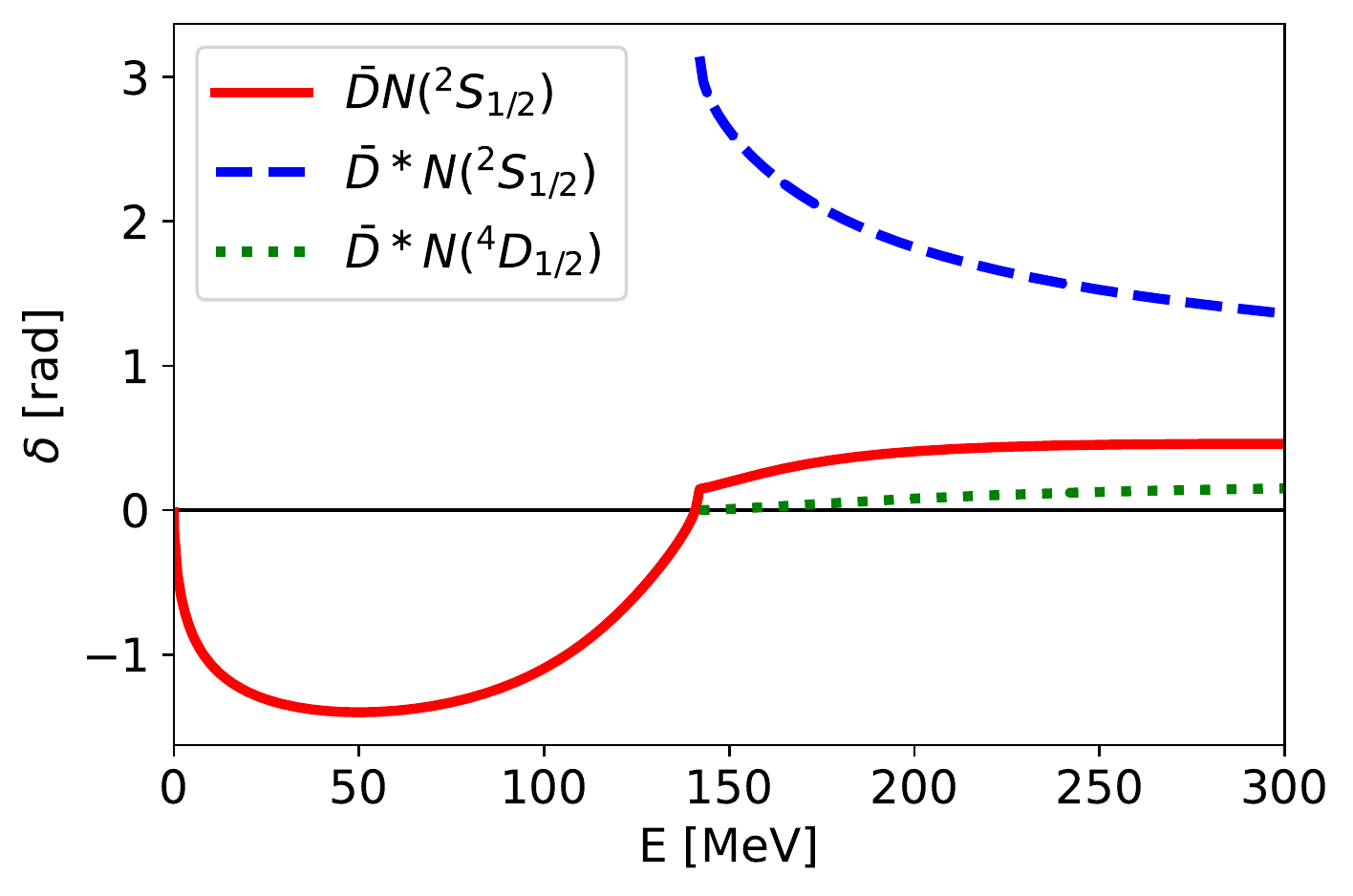} \\
 (c) $BN \, (I=0)$& (d) $BN \, (I=1)$ \\
 \includegraphics[scale=0.5]{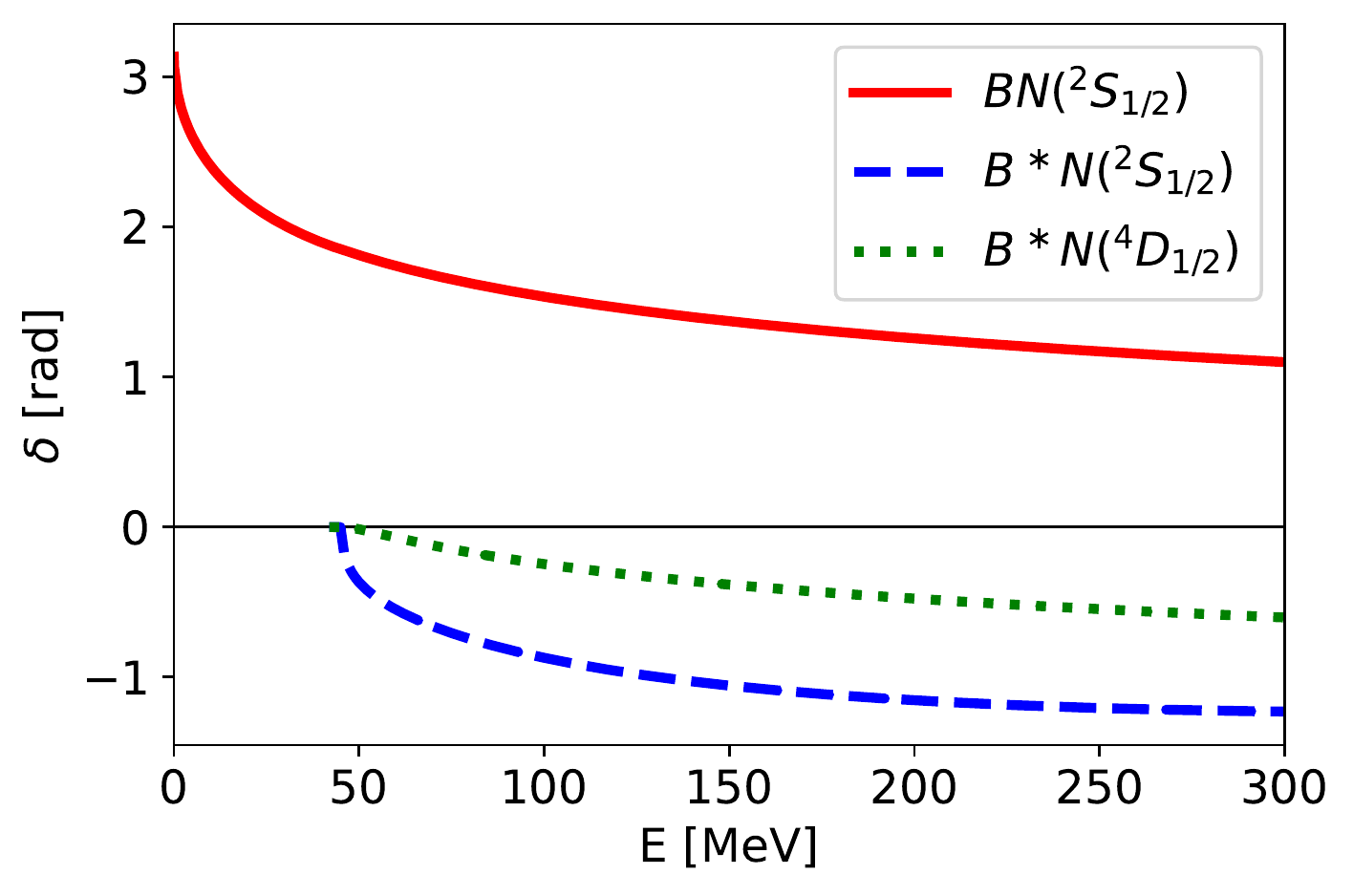} &
 \includegraphics[scale=0.5]{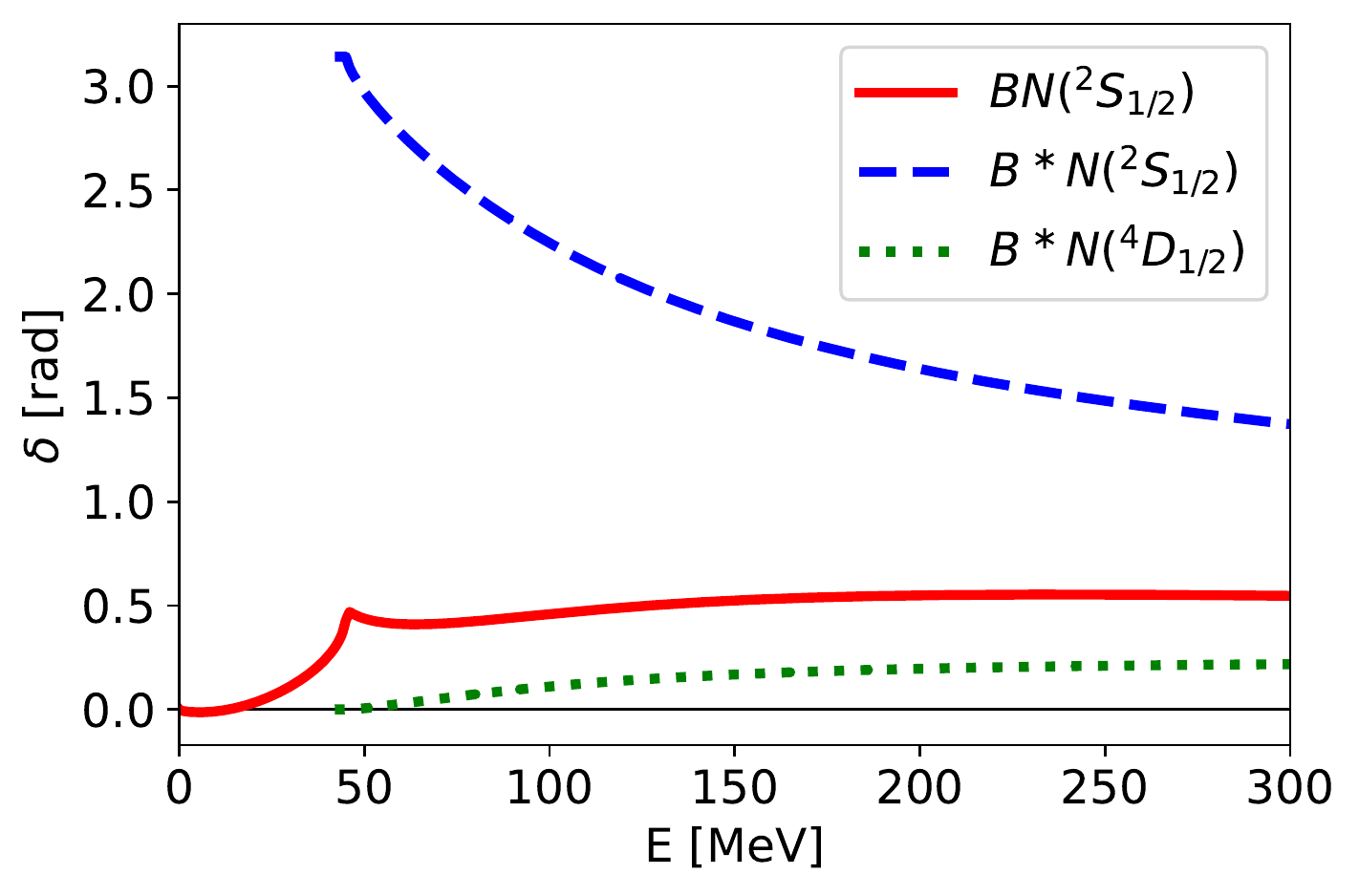} \\
\end{tabular}
\caption{The phase shifts of $\bar{D}N$ 
  [(a) and (b)]
 and $BN$ 
 [(c) and (d)]
 as functions of the scattering energy. 
 Panels
 (a) and (c)
 are for $I=0$, and 
 panels 
 (b) and (d)
 are for $I=1$. \label{fig:Phase_Shift}}
\end{center}
\end{figure*}

\renewcommand{\arraystretch}{1.25}
\begin{table}[tbp]
 \caption{Binding energies (B.E.) and mixing ratios of the $\bar{D}^{(\ast)}N$ and $B^{(\ast)}N$ states with $I(J^{P})$ quantum numbers.
 The binding energies are measured from the mass thresholds of $\bar{D}N$ or $BN$.
 \label{tbl:bound_states}}
\begin{center}  
\begin{tabular}{c|cl}
\toprule[0.3mm]
      $\bar{D}N$ & B.E. [MeV] & \hspace{1.5em} Mixing ratio [\%] \\
   \hline
      $0(1/2^-)$  & $1.38$ &  
      \begin{tabular}{ll}
         $\bar{D}N(^2{S}_{1/2})$ & $96.1$ \\ 
         $\bar{D}^\ast N(^2{S}_{1/2})$ & $1.94$ \\ 
         $\bar{D}^\ast N(^4{D}_{1/2})$ & $1.93$ \\ 
       \end{tabular} \\ 
 \hline
 $1(1/2^-)$  &  $5.99$ &
	   \begin{tabular}{cc}
	    $\bar{D}N(^2S_{1/2})$ 
	    & 88.9 \\
	    $\bar{D}^\ast N(^2S_{1/2})$ 
	    & 10.9 \\
	    $\bar{D}^\ast N(^4D_{1/2})$ 
	    & 0.11 \\
	   \end{tabular}	   \\
 \midrule[0.3mm]
 $BN$ & B.E. [MeV] & \hspace{1.5em} Mixing ratio [\%] \\ 
   \hline 
   $0(1/2^-)$ & $29.7$  &
   \begin{tabular}{ll}
      $BN(^2{S}_{1/2})$ & $76.4$ \\ 
      $B^\ast N(^2{S}_{1/2})$ & $14.1$ \\ 
      $B^\ast N(^4{D}_{1/2})$ & $9.46$ \\ 
   \end{tabular} \\ 
   \hline 
   $1(1/2^-)$ & 66.0 &
   \begin{tabular}{ll}
      $BN(^2{S}_{1/2})$ & $38.5$ \\ 
      $B^\ast N(^2{S}_{1/2})$ & $61.5$ \\ 
      $B^\ast N(^4{D}_{1/2})$ & $1.82\times 10^{-2}$ \\ 
   \end{tabular} \\ 
\bottomrule[0.3mm]
\end{tabular}
\end{center}
\end{table}
\renewcommand{\arraystretch}{1}

First let us show 
the phase shifts for $\bar{D}^{(\ast)}N$ and $B^{(\ast)}N$ scatterings with $I=0$ and $I=1$ in Fig.~\ref{fig:Phase_Shift}.
In the case of $\bar{D}N$, the $I=0$ channel has a bound state below the $\bar{D}N$ mass threshold as the phase shift starts at $\delta=\pi$ and it decreases to zero as the scattering energy increases 
 [Fig.~\ref{fig:Phase_Shift}(a)]. 
We notice that the $\bar{D}^{\ast}N$ component feels repulsion due to the existence of the 
shallow bound state.
At first sight, if we look at the phase shift of the $\bar{D}N$ component 
in the $I=1$ channel, 
then we may notice that the interaction is repulsive
and therefore no bound state exists.
However, if we turn our attention to 
the phase shift of the $\bar{D}^{\ast}N(^2S_{1/2})$ channel, 
it starts at $\delta=\pi$, indicating the presence of a bound state 
 [Fig.~\ref{fig:Phase_Shift}(b)]. 
As a result, we find a bound state that is formed 
 below
the $\bar{D}N$ threshold.
In the bottom case, the $BN$ interaction in the $I=0$ channel has a bound state below the $BN$ mass threshold, and the $B^{\ast}N$ component feels repulsion due to this bound state
 [Fig.~\ref{fig:Phase_Shift}(c)].
For $I=1$, 
the $B^\ast N(^2S_{1/2})$ phase shift also starts as $\delta=\pi$
 [Fig.~\ref{fig:Phase_Shift}(d)],  
as well as the $\bar{D}^\ast N(^2S_{1/2})$ one, 
indicating that there is a bound state driven by the $B^\ast N$ component.

In Table~\ref{tbl:bound_states}, 
we summarize the binding energies and the mixing ratios of 
$PN$ and $P^\ast N$ components.
The bound $\bar{D}N$ state in $I=0$ has the binding energy 1.38 MeV.
The 
state is almost dominated by $\bar{D}N(^{2}{S}_{1/2})$ with a small mixture of $\bar{D}^{\ast}N(^{2}{S}_{1/2})$ and $\bar{D}^{\ast}N(^{4}{D}_{1/2})$.
Even when the 
amount of
$D$-wave component is 
small,
it plays an important role to 
provide
attraction by the tensor interaction in the OPEP as emphasized in our previous papers~\cite{Yasui:2009bz,Yamaguchi:2011xb,Yamaguchi:2011qw}.
In $I=1$, we also obtain the bound state with the binding energy $5.99$ MeV.
In contrast to the isosinglet state, the bound state has a few amount of 
the $\bar{D}^{\ast}N(^{4}{D}_{1/2})$ component.
This suggests that the $\bar{D}N$ bound state with $I=1$ is generated mainly not by the OPEP but by 
the other potentials. 
In the present model setting, in fact, the $\sigma$ exchange potential provides a strong attraction in the $P^{(\ast)}N$ systems as the $\sigma$ exchange potential is strongly attractive for the $NN$ system with $I=1$ in the CD-Bonn potential.
In the bottom case, the $BN$ states with $I=0$ and $I=1$ give deeply bound states with the binding energies 
 29.7 
and 66.0 MeV, respectively.
In $I=0$, the main component is 
$BN(^{2}{S}_{1/2})$ 
with a 
small amount of 
 $B^{\ast}N(^{2}{S}_{1/2})$ and $B^{\ast}N(^{4}{D}_{1/2})$
components.
The existence of the $D$-wave component indicates again the importance of the OPEP.
In $I=1$, the $D$-wave component is negligible as seen in the $\bar{D}N$ bound state. Interestingly, the $B^\ast N(^2S_{1/2})$ channel dominates in the isotriplet bound state, which will be discussed in Sec.~\ref{sec:discussion}.
The scattering lengths in each state are summarized in Table~\ref{tbl:scattering_lengths}.

The phase shifts for $I=1$ in Figs.~\ref{fig:Phase_Shift}(b) and \ref{fig:Phase_Shift}(d), starting at $\delta=\pi$, imply the existence of the $\bar{D}^\ast N$ and $B^\ast N$ bound states.
{ 
In order to confirm this idea, 
we have performed
}
a bound-state analysis considering only the $\bar{D}^\ast N$ ($B^\ast N$) channels, 
when the $\bar{D}N$ ($BN$) channel is switched off. 
As a result, 
we find $\bar{D}^\ast N$ and $B^\ast N$ bound states with the binding energies 
 $29.1$
and $51.0$ MeV, respectively,
measured from the $\bar{D}^\ast N$ ($B^\ast N$) threshold.

\begin{figure*}[t]
\begin{center}
\begin{tabular}{cc}
 (a) $\bar{D}N\, (I=0)$& (b) $\bar{D}N \, (I=1)$ \\
 \includegraphics[scale=0.5]{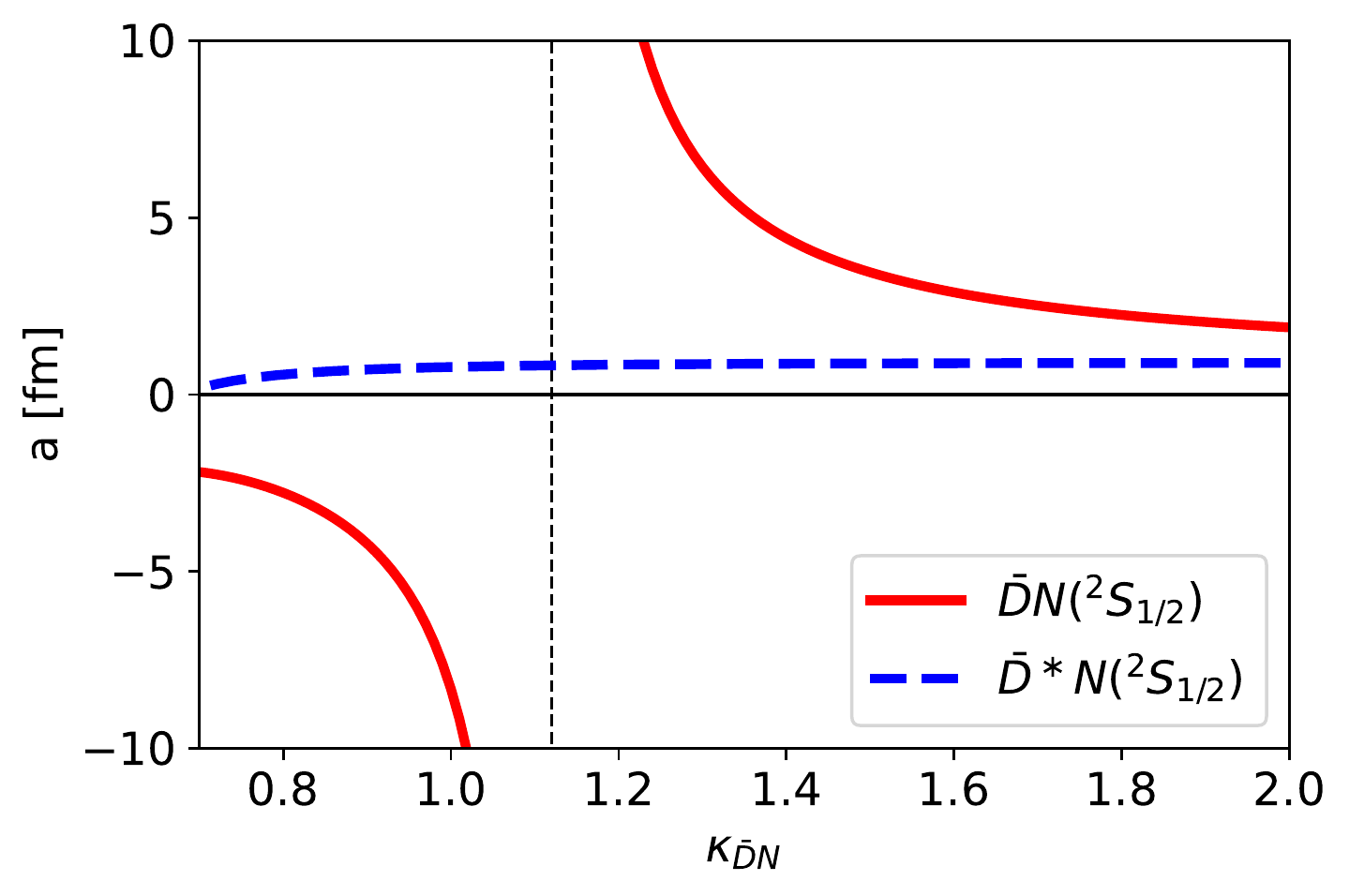} & 
     \includegraphics[scale=0.5]{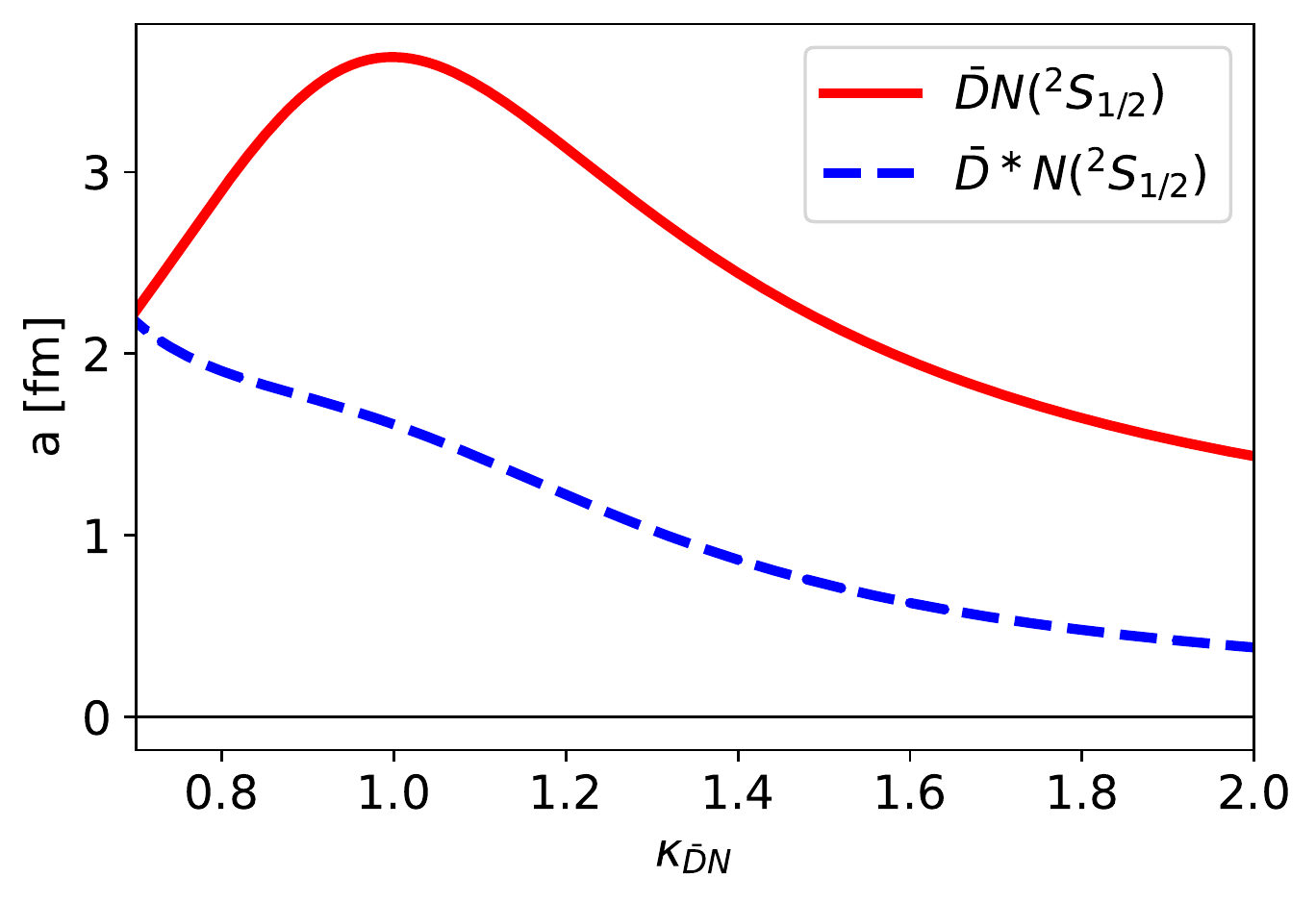} \\ 
 (c) $BN \, (I=0)$& (d) $BN \, (I=1)$ \\
 \includegraphics[scale=0.5]{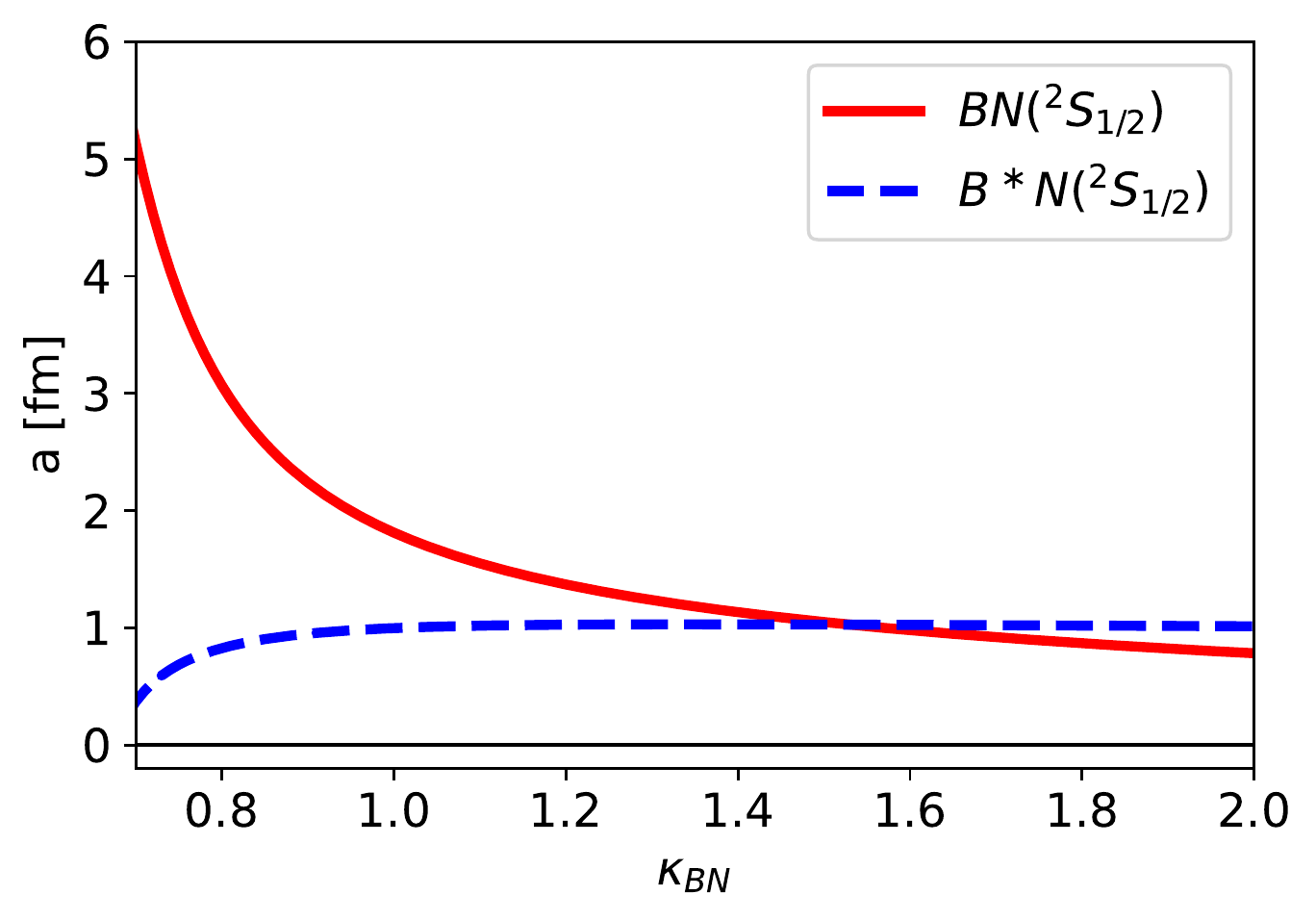} & 
\includegraphics[scale=0.5]{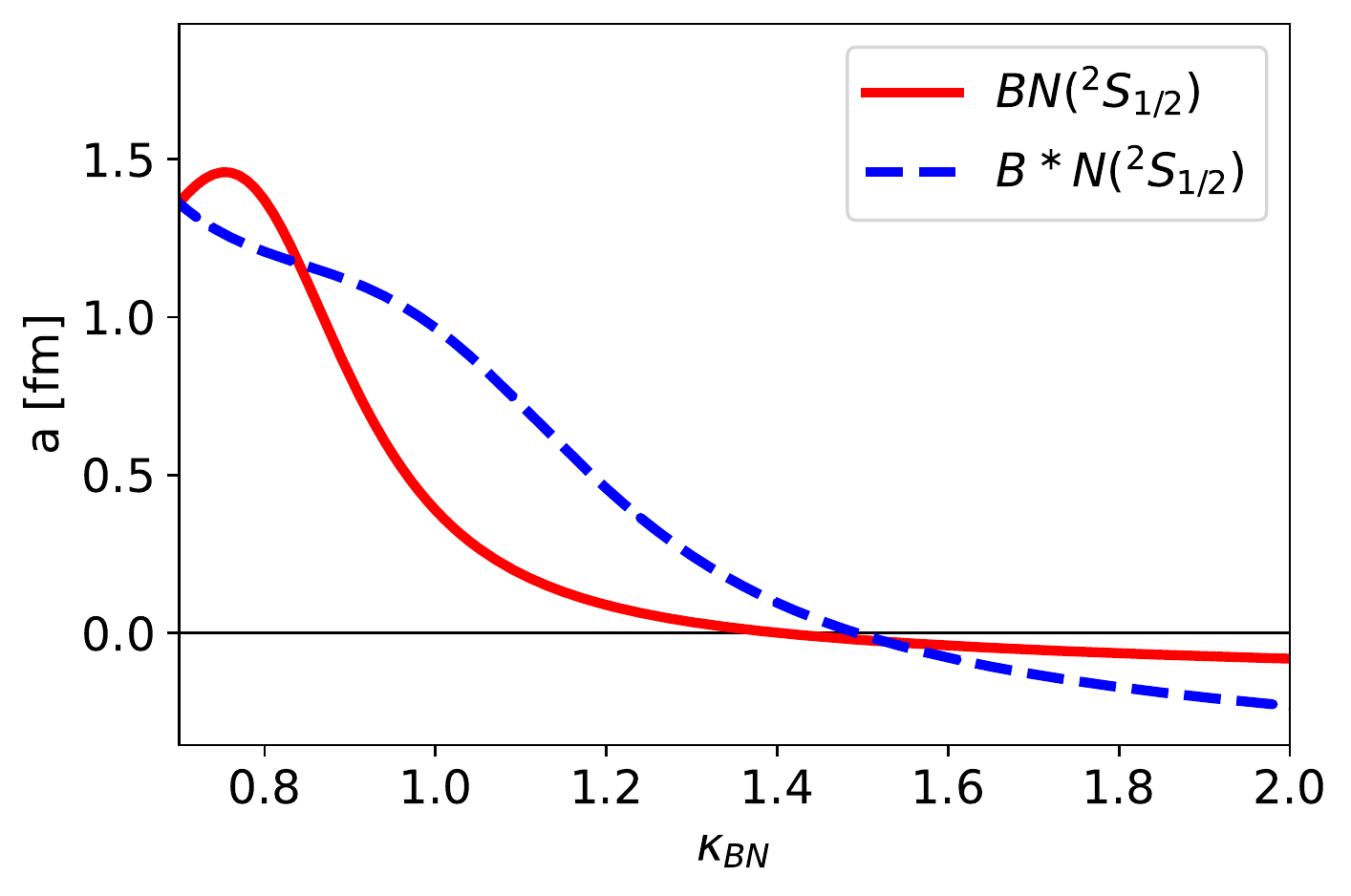} \\
\end{tabular}
\caption{The scattering lengths of $\bar{D}N$ 
 [(a) and (b)]
 and 
 $BN$ 
 [(c) and (d)]
 as functions of the cutoff ratio $\kappa_{\bar{D}N}$ and $\kappa_{BN}$. 
Panels 
 (a) and (c)
 are for $I=0$, and 
 panels 
 (b) and (d)
 are for $I=1$. \label{fig:a_vs_Cutoff}}
\end{center}
\end{figure*}

We investigate the parameter dependence of the attraction in $P^{(\ast)}N$,
where the values of these parameters have some ambiguity in the present model setting.
In Fig.~\ref{fig:a_vs_Cutoff}, we show the dependence of the scattering lengths on the cutoff-ratio parameters, $\kappa_{\bar{D}N}$ and $\kappa_{BN}$.
In the $\bar{D}N$ case, we find that the attraction in $I=0$ is provided for $\kappa_{\bar{D}N} \simge 1.1$ 
whose
values are consistent with the one estimated by the ratio of the different hadron sizes of a $\bar{D}$ meson and a nucleon, as previously discussed in Refs.~\cite{Yasui:2009bz,Yamaguchi:2011xb,Yamaguchi:2011qw}.
The 
strength of attraction
in $I=1$ is 
not so dependent on the choice of $\kappa_{\bar{D}N}$.
In the $BN$ case, the attraction in $I=0$ has only weak dependence on the choice of $\kappa_{BN}$ in the range of $\kappa_{BN} \simge 1.0$.
This result would tell us a confidence for the existence of the $BN$ bound state in $I=0$.
In comparison with $I=0$, the attraction in $I=1$ is more sensitive to choice of the value of $\kappa_{BN}$.
Thus the 
{ deeply $BN$ } 
bound state in $I=1$ needs to be carefully considered in terms of its model dependence.

\begin{figure*}[htbp]
 \begin{center}  
 \begin{tabular}{ccc}
  (a) Binding energy& (b) Scattering length for $PN$& 
	  (c) Scattering length for $P^\ast N$ \\
  \includegraphics[width=0.33\linewidth]{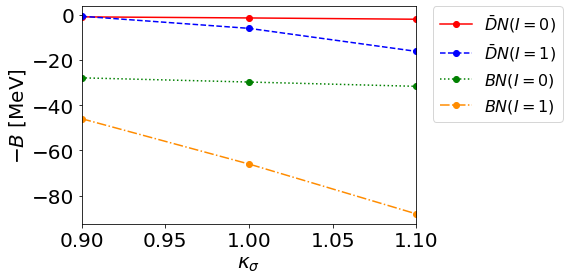}& 
      \includegraphics[width=0.33\linewidth]{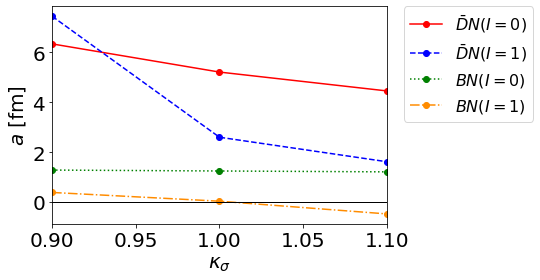} & 
	  \includegraphics[width=0.33\linewidth]{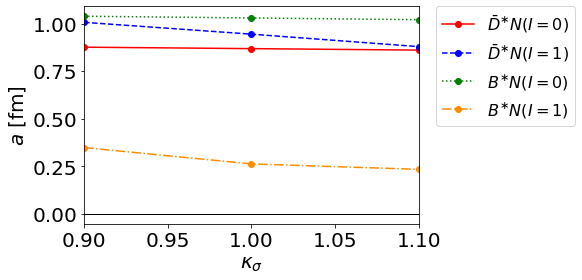} \\
\end{tabular}
  \caption{\label{fig:sigma_dependence}   
  The dependence of 
  the binding energies [panel (a)] and scattering lengths of the $PN$ and $P^\ast N$ states [panels (b) and (c)] 
  on the sigma coupling strengths.
  The scale parameter $\kappa_\sigma$ is introduced as $g^\prime_\sigma=\kappa_\sigma g_{\sigma}$, 
  i.e., changing $g_\sigma$ to $g^\prime_\sigma$ as a free coupling value.      
  }
 \end{center}
\end{figure*}

Uncertainty in the current model is also brought by the sigma coupling.
In general,
 the
coupling constants of the meson exchange potential are fixed by the experimental data, such as 
the nucleon-nucleon scattering data and heavy meson decays. 
However, the sigma coupling to 
the heavy meson 
is difficult to be determined uniquely only by the currently existing experimental data. In our present 
calculation
framework, we have adopted $g_\sigma = g_{\sigma NN}/3$ (see Sec.~II A2).
In order to investigate the uncertainty from the ambiguity of the sigma coupling value,
we estimate 
{ the }
dependence of 
binding energies and scattering lengths on the $g_\sigma$ coupling constant as shown in Fig~\ref{fig:sigma_dependence}.
Here we show (a) the binding energies, (b) the scattering lengths for $PN$, and (c) the scattering length for $P^\ast N$, respectively, for $\bar{D}N (I=0)$, $\bar{D}N (I=1)$, $BN (I=0)$, and $BN (I=1)$.
For $I=0$, the binding energies and the scattering lengths are 
not sensitive to $g_\sigma$ 
value indicating 
that the sigma exchange force is not dominant in $I=0$; the pion exchange is the most dominant.
For $I=1$, however, 
the results indicate the 
sensitiveness to
$g_\sigma$, i.e., that the sigma exchange force is dominant rather than the other meson exchanges.

The existence of the $\bar{D}N$ and $BN$ bound states in $I=0$ is consistent with the result in our previous works~\cite{Yasui:2009bz,Yamaguchi:2011xb,Yamaguchi:2011qw}.
However, we should 
note
the difference between the present analysis and the previous one.
In the previous case, the $\pi$ exchange potential was almost dominant among the $\pi$, $\rho$, and $\omega$ exchanges.
However, 
the coupling strengths 
{ of the meson exchange potentials}
were 
incorrectly overestimated 
by 
 a factor of 2
due to the incorrect normalization of wave functions
in Refs.~\cite{Yasui:2009bz,Yamaguchi:2011xb,Yamaguchi:2011qw}.
In the present 
analysis for $I=0$, 
we have also found that similar bound states exist
by reconstructing the $PN$ interaction model newly including the $\sigma$ exchange.
Again, the $\pi$ exchange potential plays the dominate role to produce the attraction.
In contrast, the bound states in $I=1$ have been obtained in the $PN$ states, where the main attraction is provided by the $\sigma$ potential 
whose strength in $I=1$ is set to be larger than that in $I=0$.

\section{Discussion} \label{sec:discussion}

We discuss the internal spin structures of the bound $\bar{D}N$ and $BN$ states in a view of the HQS 
symmetry.
As already discussed 
in detail in Ref.~\cite{Yamaguchi:2014era}, the $P^{(\ast)}N$ state can be decomposed 
into 
product states of the heavy antiquark $\bar{Q}$ and the light quarks $qqqq$ in the heavy quark limit.
The latter component is called the 
 {\it light spin complex}, 
instead of the brown muck, because 
it makes a 
 specific structure composed 
of $q$ and $N$ which is denoted by $[qN]_{j^{\cal P}}$ with total spin $j$ and parity ${\cal P}$ of 
the light quark components.
These are a conserved 
quantities
due to the spin decoupling from the heavy quark.
The important property in the heavy quark limit is that the ratio of the fractions of the amount of $PN(^{2S+1}L_{J})$ and $P^{\ast}N(^{2S'+1}L'_{J})$ 
 wave functions 
is determined uniquely. Here $S'$ and $L'$ can be different from $S$ and $L$, respectively, in general.
As shown explicitly in Ref.~\cite{Yamaguchi:2014era}, we obtain the fractions
\begin{align}
   PN(^{2}{S}_{1/2}):P^{\ast}N(^{2}{S}_{1/2})=1:3,
\label{eq:ration_0}
\end{align}
for $j^{\cal P}=0^{+}$ and
\begin{align}
   PN(^{2}{S}_{1/2}):P^{\ast}N(^{2}{S}_{1/2})=3:1,
\label{eq:ration_1}
\end{align}
for $j^{\cal P}=1^{+}$,
which hold 
irrespectively  
of the choice of the $PN$-$P^{\ast}N$ potential.
Although these ratios are exact only in the heavy quark limit, they provide us with a guideline to understand the internal spin structures of the obtained $\bar{D}N$ and $BN$ bound states.

In Table~\ref{tbl:bound_states}, for example, we 
show
that the mixing ratios of $BN(^{2}{S}_{1/2})$ and $B^{\ast}N(^{2}{S}_{1/2})$ in $I=0$ are 76.4 \% and 14.4\%, respectively, 
which
are close to the ratio in 
Eq.~\eqref{eq:ration_1} 
rather than that in 
Eq.~\eqref{eq:ration_0}. 
Thus, it is suggested that the $BN$ bound state in $I=0$ 
is dominated by
the light 
 spin complex
with $j^{\cal P}=1^{+}$.
In contrast, 
the mixing ratios  $BN(^{2}{S}_{1/2})$ and $B^{\ast}N(^{2}{S}_{1/2})$ in $I=1$ are 38.5 \% and 61.5 \%, respectively, are close to the ratio in 
Eq.~\eqref{eq:ration_0} 
rather than that in 
Eq.~\eqref{eq:ration_1}. 
Thus, it is suggested that the $BN$ bound state in $I=1$ includes the light 
spin complex
with 
$j^{\cal P}=0^{+}$
as a major component.

One may wonder 
that the ratios in bottom sector are not the same as the ratios in Eqs.~\eqref{eq:ration_0} and \eqref{eq:ration_1} in spite of the sufficient heaviness of the bottom quark mass.
This would be simply due to the violation of the heavy quark spin symmetry stemming from the difference of the $B$ meson mass and the $B^{\ast}$ meson mass, as noted in Ref.~\cite{Yamaguchi:2014era}.

We should notice that the existence of the $j^{\cal P}=0^{+}$ state is new because only the $j^{\cal P}=1^{+}$ state was reported for the $\pi$, $\rho$, and $\omega$ potentials in Ref.~\cite{Yamaguchi:2014era}.
We can understand this new result in terms of 
the fact
that the $j^{\cal P}=0^{+}$ state is provided mainly by the $\sigma$ potential because of the sufficient attraction in the $\sigma_{1}$ exchange stemming from the characteristic property of the CD-Bonn potential (see 
 Table~\ref{tbl:NN_parameter} in Appendix~\ref{sec:NN_potential}).

\renewcommand{\arraystretch}{1.25}
\begin{table}[tbp]
 \caption{$S$-wave scattering lengths ($a$) of the $\bar{D}^{(\ast)}N$ and $B^{(\ast)}N$ states.
 An attractive scattering length is given by the negative sign ($a < 0$), and a repulsive scattering length and the scattering length for a bound state are given by the positive sign ($a>0$).
\label{tbl:scattering_lengths}}
\begin{center}  
\begin{tabular}{c|l}
\toprule[0.3mm]
   $\bar{D}N$ & \hspace{4em} $a$ [fm] \\ 
   \hline
   $0(1/2^-)$ &
   \begin{tabular}{ll}
      $\bar{D}N(^2{S}_{1/2})$ & 5.21 \\ 
      $\bar{D}^\ast N(^2{S}_{1/2})$ & $0.868 - i 3.72\times 10^{-2}$ \\ 
   \end{tabular} \\ 
   \hline
   $1(1/2^-)$ &
   \begin{tabular}{ll}
      $\bar{D}N(^2{S}_{1/2})$ & 2.60 \\ 
      $\bar{D}^\ast N(^2{S}_{1/2})$ & $0.944 - i 0.722$ \\ 
   \end{tabular} \\ 
   \midrule[0.3mm]
   $BN$ & \hspace{4em} $a$ [fm] \\ 
   \hline
   $0(1/2^-)$ &
   \begin{tabular}{ll}
      $BN(^2{S}_{1/2})$ & 1.25 \\ 
      $B^\ast N(^2{S}_{1/2})$ & $1.03 - i 1.07 \times 10^{-2}$ \\ 
   \end{tabular} \\ 
   \hline
   $1(1/2^-)$ &
   \begin{tabular}{ll}
      $BN(^2{S}_{1/2})$ & $3.84 \times 10^{-2}$ \\ 
      $B^\ast N(^2{S}_{1/2})$ & $0.263-i0.585$ \\ 
   \end{tabular} \\ 
\bottomrule[0.3mm]
\end{tabular}
\end{center}
\end{table}
\renewcommand{\arraystretch}{1}

\section{Conclusion} \label{sec:conclusion}

We have discussed the $\bar{D}^{(\ast)}N$ and $B^{(\ast)}N$ bound states in terms of the $\pi$, $\sigma$, $\rho$, and $\omega$ meson-exchange potentials by considering the heavy-quark spin symmetry and the chiral symmetry.
By referring the CD-Bonn potential for the nuclear force, we have constructed the $PN$-$P^{\ast}N$ potential with the $\sigma$ exchanges as new degrees of freedom at middle-range interaction.
We have carefully calculated the potentials with appropriate factors stemming from the normalization of the wave function which were underestimated in our previous studies~\cite{Yasui:2009bz,Yamaguchi:2011xb,Yamaguchi:2011qw}.
As results, we have found 
that the interaction is largely attractive to hold 
the $\bar{D}N$ bound state and the $BN$ bound state below the lowest mass threshold for each in $I(J^{P})=0(1/2^{-})$ channel.
Their binding energies are close to the values 
that
were obtained by our previous works.
With the present potential including $\sigma$ exchange, interestingly, we have found that the $\sigma$ exchange as well as the $\pi$ exchange still plays an important role.
We also have found 
the $\bar{D}N$ and $BN$ 
bound states 
in $I(J^{P})=1(1/2^{-})$ as a new state which has not been discussed so far.
It is expected that those states are relevant to the $D^{-}p$ interaction researched in LHCb~\cite{alicecollaboration2022study}.

The attraction in $PN$-$P^{\ast}N$ systems would open a new way to understand the inter-hadron interaction in heavy flavors.
It is important that these systems are made of genuinely five-quark components due to the absence of the annihilation channels.
It may help us to understand the new channels of exotic hadrons.
Furthermore, the many-body dynamics would be an interesting subject, because the $PN$-$P^{\ast}N$ attraction suggests the formation of heavy-flavored nuclei as 
many-body states having the 
impurity particles in nuclei~\cite{Hosaka:2016ypm}. 
Few-body systems such as $\bar{D}NN$ ($BNN$)~\cite{Yamaguchi:2013hsa} and $\bar{D}\alpha$ ($^{\bar{D}}$He) 
($B\alpha$ ($^{B}$He))
are also interesting, which can be accessed through the relativistic heavy ion collisions in LHC and RHIC~\cite{ExHIC:2010gcb,ExHIC:2011say,ExHIC:2017smd}. 
The nuclear structure of charm and bottom nuclei has been studied theoretically for some possible exotic light nuclei ~\cite{Yamaguchi2017mn}.
Experiments at J-PARC, GSI-FAIR, NICA, and so on would also be interesting.
In theoretical study, the cross sections for producing  such exotic nuclei
have been discussed~\cite{Yamagata2016fs}.
As one of the advanced topics related to heavy-flavored nuclei, the isospin Kondo effect is interesting as it exhibits the ``confinement" of isospin charge~\cite{Yasui2013hq,Yasui2016ke,Yasui2017ke,Yasui2017kc,Yasui2019si}.
Many subjects are 
awaiting to be discussed in the future.

\section*{Acknowledgment}
This work is in part supported by 
Grants-in-Aid for Scientific Research under Grant Numbers JP20K14478 (Y.Y.) and 21H04478(A) (A.H.).
A.H. was also supported by Grant-in-Aid for Scientific Research on Innovative Areas (No. 18H05407).

\appendix

\section{The $NN$ potential} \label{sec:NN_potential}

We construct the nuclear potential
by considering the $\pi$, $\sigma$, $\rho$, and $\omega$ exchanges.
Their interaction Lagrangians for the vertices with a nucleon are given by
\begin{align}
   {\cal L}_{\pi NN} = \, &
 g_{\pi NN} 
 \bar{\psi} i\gamma_{5} \vec{\tau}\!\cdot\vec{\pi} \psi, 
 \label{eq:Lpi_NN}
 \\ 
   {\cal L}_{\sigma_{I} NN} = \, &
 g_{\sigma_{I} NN}  
 \bar{\psi}\sigma_{I}\psi, \\ 
   {\cal L}_{\rho NN} = \, &
  g_{\rho NN} 
 \bar{\psi} \gamma_{\mu} \vec{\tau}\!\cdot\!\vec{\rho}^{\mu} \psi
 \notag\\
  & + 
 \frac{f_{\rho NN}}{4m_{N}} \bar{\psi} \sigma_{\mu\nu} \vec{\tau}\!\cdot\! \bigl( \partial^{\mu} \vec{\rho}^{\nu} - \partial^{\nu} \vec{\rho}^{\mu} \bigr) \psi, \\ 
   {\cal L}_{\omega NN} = \, &
  g_{\omega NN} 
 \bar{\psi} \gamma_{\mu}\omega^{\mu} \psi,
 \label{eq:Lomega_NN}
\end{align}
with the appropriate coupling constants.
We use different $\sigma$ mesons: the $\sigma_{0}$ meson 
for
the isosinglet ($I=0$) $NN$ scatterings and the $\sigma_{1}$ meson 
for
the isotriplet ($I=1$) $NN$ scatterings.
Their difference appears not only in the coupling constants but also in their masses.
We sometimes omit the underscript $I$ if unnecessary.
From the Lagrangians~\eqref{eq:Lpi_NN}-\eqref{eq:Lomega_NN},
we obtain the $NN$ potentials:
\begin{align}
   V_{\pi}(r)
= \, &
   \biggl(\frac{g_{\pi NN}}{2m_{N}}\biggr)^{2}
   \frac{1}{3}
   \Bigl(
         \vec{\sigma}_{1}\!\cdot\!\vec{\sigma}_{2} C_{\pi}(r)
      + S_{12}(\hat{\vec{r}}) T_{\pi}(r)
   \Bigr)
   \vec{\tau}_{1}\!\cdot\!\vec{\tau}_{2}
\nonumber \\ 
\equiv \, &
 \Bigl(
   \vec{\sigma}_{1}\!\cdot\!\vec{\sigma}_{2} C_{\pi}^{NN}(r)
+ S_{12}(\hat{\vec{r}}) T_{\pi}^{NN}(r)
 \Bigr)
 \vec{\tau}_{1}\!\cdot\!\vec{\tau}_{2},
\\ 
   V_{v}(r)
= \, &
   g_{vNN}^{2}
   \biggl(\frac{1}{m_{v}^{2}}+\frac{1+f_{vNN}/g_{vNN}}{2m_{N}^{2}}\biggr)
   C_{v}(r)
   \nonumber \\ & 
+ g_{vNN}^{2}
   \biggl(\frac{1+f_{vNN}/g_{vNN}}{2m_{N}}\biggr)^{2}
   \nonumber \\ & \times 
   \frac{1}{3}
   \Bigl(
         2\vec{\sigma}_{1}\!\cdot\!\vec{\sigma}_{2} C_{v}(r)
       - S_{12}(\hat{\vec{r}}) T_{v}(r)
   \Bigr)
\nonumber \\ 
\equiv \, &
   C_{v}^{\prime NN}(r)
+ 2\vec{\sigma}_{1}\!\cdot\!\vec{\sigma}_{2} C_{v}^{NN}(r)
 - S_{12}(\hat{\vec{r}}) T_{v}^{NN}(r),
\\ 
   V_{\sigma_{I}}(r)
= \, &
 - \biggl(\frac{
  g_{\sigma_{I} NN} 
 }{2m_{N}}\biggr)^{2}
   \Biggl(\biggl(\frac{2m_{N}}{m_{\sigma_{I}}}\biggr)^{2}-1\Biggr)
   C_{\sigma_{I}}(r)
\nonumber \\ 
\equiv \, &
 - C_{\sigma_{I}}^{NN}(r),
\end{align}
with $v=\rho$, $\omega$, where the functions $C_{\pi}^{NN}$, $T_{\pi}^{NN}$, $C_{\sigma_{I}}^{NN}$,  $C_{v}^{\prime NN}$, $C_{v}^{NN}$, and $T_{v}^{NN}$ are defined as above.
More concretely, the $NN$ potentials are expressed by
\begin{align}
   V_{^{3}{S}_{1}}^{NN}(r)
&=
   \bar{V}_{\pi}^{NN}(r) + \bar{V}_{\sigma_{0}}^{NN}(r) + \bar{V}_{\rho}^{NN}(r) + \bar{V}_{\omega}^{NN}(r),
\end{align}
with
\begin{align}
   \bar{V}_{\pi}^{NN}(r)
&=
   \left(
   \begin{array}{cc}
    -3C_{\pi}^{NN} & -6\sqrt{2} T_{\pi}^{NN} \\
    -6\sqrt{2} T_{\pi}^{NN} & -3C_{\pi}^{NN}+6T_{\pi}^{NN}
   \end{array}
   \right), \\
   \bar{V}_{v}^{NN}(r)
&=
   \left(
   \begin{array}{cc}
    C_{v}^{\prime NN}+2C_{v}^{NN} & -2\sqrt{2} T_{\pi}^{NN} \\
    -2\sqrt{2} T_{\pi}^{NN} & C_{v}^{\prime NN}+2C_{v}^{NN}+2T_{v}^{NN}
   \end{array}
   \right), \\
   \bar{V}_{\sigma_{0}}^{NN}
&=
   \left(
   \begin{array}{cc}
    -C_{\sigma_{0}}^{NN} & 0 \\
    0 & -C_{\sigma_{0}}^{NN}
   \end{array}
   \right),
\end{align}
in the $^{3}{S}_{1}$ channel, where the $^3S_1$ and $^3D_1$ components are coupled,
and
\begin{align}
   V_{^{1}{S}_{0}}^{NN}(r)
&=
   V_{\pi}^{NN}(r) + V_{\sigma_{1}}^{NN}(r) + V_{\rho}^{NN}(r) + V_{\omega}^{NN}(r),
\end{align}
with
\begin{align}
   V_{\pi}^{NN}(r) &= -3C_{\pi}^{NN}(r), \\ 
   V_{v}^{NN}(r) &= C_{v}^{\prime NN}(r) -6C_{v}^{NN}, \\ 
   V_{\sigma_{1}}^{NN} &= -C_{\sigma_{1}}^{NN},
\end{align}
in the $^{1}{S}_{0}$ channel.
Notice that the tensor potentials are switched on due to the spin-1 property in the $I=0$ channel.

We choose the values of the coupling constants to be the same values as those in the CD-Bonn potential~\cite{Machleidt:2000ge} as summarized in Table~\ref{tbl:NN_parameter}.
We notice that the CD-Bonn potential 
originally includes 
the nonlocal potentials in the $\pi$, $\sigma$, $\rho$, and $\omega$ exchanges, and contact terms stemming from the short-range part in the meson exchange.
In the present study, however, we neglect the nonlocal potentials, the contact terms and massive $\sigma$ mesons, and so on, because we are interested only in the low-energy parts in the $NN$ scatterings.

In order to compensate the difference from the CD-Bonn potential, we rescale the 
cutoff parameter by introducing 
$\kappa_{I}$ providing 
the new cutoffs 
$\Lambda_N = \kappa_{I}\Lambda_{N}^{\text{CD-Bonn}}$. 
Here 
$\Lambda_{N}^{\text{CD-Bonn}}$ 
is the original cutoff parameter in the CD-Bonn potential~\cite{Machleidt:2000ge}, 
whose values 
depend on the exchanged mesons, 
$\pi$, $\sigma$, $\rho$, and $\omega$.
$\kappa_I$ ($I=0$ and $I=1$) are the scale parameter, introduced newly for the adjustment to reproduce the low-energy $NN$ scatterings in the present simple model of nuclear force.
Notice the values of $\kappa_{I}$ are dependent only on the isospin channels $I=0$ and $I=1$,
while they are common to the $\pi$, $\sigma$, $\rho$, and $\omega$ exchanges.
We use the values in proton-neutron channel in $I=1$ in the CD-Bonn potential, because the electric Coulomb force is not included in our potential.
We determine the values of $\kappa_{I}$ to reproduce the binding energy of a deuteron $B_{\text{d}}$ in the $^{3}{S}_{1}$ ($I=0$) channel as well as the $NN$ scattering length in the $^{1}{S}_{0}$ ($I=1$) channel.
As the best fitting, we obtain $\kappa_{0}=0.804$ for $I=0$ and $\kappa_{1}=0.773$ for $I=1$.
Roughly, we consider that those values would represent the ``effective" cutoff parameters when the higher-energy dynamics is renormalized at lower energy near thresholds.
Similar values are obtained also when the $NN$ scattering length in the $^{3}{S}_{1}$ ($I=0$) channel is chosen instead of $B_{\text{d}}$.
As shown in 
Table~\ref{tbl:NN_scattering_length},  
the obtained values of the scattering lengths and the effective ranges are well consistent with those obtained from the original CD-Bonn potential,
$a(^{3}{S}_{1})=5.419\pm0.007$ fm, $r_{\text{e}}(^{3}{S}_{1})=1.753\pm0.008$ fm,
$a(^{1}{S}_{0})=-23.740\pm0.020$ fm, $r_{\text{e}}(^{1}{S}_{0})=2.77\pm0.05$ fm,
and $B_{\text{d}}=2.225$ MeV,
see Ref.~\cite{Machleidt:2000ge} for details.

\begin{table}[t] 
\caption{\label{tbl:NN_parameter}  Parameters of the local $NN$ potentials. 
 $\sigma_{I}$ is the $\sigma$ meson considered in the $NN$ scatterings 
 for the isosinglet $(I=0)$ and isotriplet $(I=1)$ channels. 
 The meson masses are given as the isospin-averaged values.  
 The coupling constants are taken from Ref.~\cite{Machleidt:2000ge}. 
 The cutoff parameters $\Lambda_N$ are obtained by scaling the original cutoffs in the CD-Bonn potential~\cite{Machleidt:2000ge}
 by the 
 parameter $\kappa_I$, where $\kappa_0=0.804$ and $\kappa_1= 0.773$ (cf. Table~\ref{tbl:NN_scattering_length}), see details in text. 
 }
\begin{center} 
\begin{tabular}{cccccc}
\hline
   \multirow{2}{*}{Mesons ($\alpha$)} & \multirow{2}{*}{Masses [MeV]} & 
 \multirow{2}{*}{$\frac{g_{\alpha NN}^{2}}{4\pi}$}  
 & \multirow{2}{*}{$\frac{f_{\alpha NN}}{g_{\alpha NN}}$} 
 & \multicolumn{2}{c}{$\Lambda_N$ [MeV]}
		 \\
 & & & &$I=0$ &$I=1$ \\ 
\hline
   $\pi$ & 138.04 & 13.6 & --- 
	     & 
		 1384 & 1330
		 \\
   $\rho$ & 769.68 & 0.84 & 6.1 
	     & 
		 1054 & 1013
		 \\
   $\omega$ & 781.94 & 20 & 0.0 
	     & 
		 1207 & 1159
		 \\
   $\sigma_{0}$ & 350 & 0.51673 & --- 
	     	     & 
		 2011 & ---
		 \\
   $\sigma_{1}$ & 452 & 3.96451 & --- 
	     & 
		 --- & 1932
	     \\
\hline
\end{tabular}
\end{center}
\end{table}

\begin{table}[t] 
\caption{The scale parameters $\kappa_{I}$ ($I=0$ and $I=1$) and the observables in the $NN$ scatterings. $a$ and $r_{e}$ are the  scattering length and the effective range, respectively. $B_{\text{d}}$ is the binding energy of a deuteron in $I=0$. The values with * indicate the input values. \label{tbl:NN_scattering_length}}
\begin{center}
\begin{tabular}{ccccc}
\hline
 { Channel} & $\kappa_{I}$ $(I=0,1)$ 
     & $a$ [fm] & $r_{\text{e}}$ [fm] & $B_{\text{d}}$ [MeV] \\
\hline
 $^{3}{S}_{1}$ ($I=0$) & 0.804   
& 5.296 & 1.562 & 2.225* \\
 $^{1}{S}_{0}$ ($I=1$) & 0.773   
& $-23.740$*
	 & 2.337 & --- \\ 
\hline
\end{tabular}
\end{center}
\end{table}

\section{Potential in a simple model} \label{sec:deriving_potential}

As an illustration of deriving a potential, we consider a simple model where a potential is provided by the boson exchange interaction ($\phi$) between two heavy particles ($\Phi$).
We consider the Lagrangian
\begin{align}
   {\cal L}[\phi,\Phi]
= \, &
   \frac{1}{2}
   \bigl( \partial_{\mu}\phi \, \partial^{\mu} \phi - m^{2} \phi^{2} \bigr)
 - g \phi \Phi^{\dag}\Phi
   \nonumber \\ & 
+ \partial_{\mu}\Phi^{\dag} \, \partial^{\mu} \Phi - M^{2} \Phi^{\dag} \Phi,
\label{eq:Lagrangian_simple}
\end{align}
with the masses $m$ and $M$ for $\phi$ and $\Phi$, respectively.
From the equation of motion for $\phi$,
$
   (\partial^{2}+m^{2})\phi
=
 - g \Phi^{\dag}\Phi
$,
we obtain the solution
\begin{align}
   \phi(x)
=
   g
   \int \dr^{4}y \,
   \langle x |
   \biggl( \frac{-1}{\partial^{2}+m^{2}} \biggr)_{xy}
   | y \rangle
   \Phi^{\dag}(y)\Phi(y),
\end{align}
for given $\Phi(y)$.
As a nonrelativistic limit,
making the approximation $\partial^{2} = \partial_{0}^{2}-\vec{\partial}^{2} \approx -\vec{\partial}^{2}$,
we find that the solution is expressed by
\begin{align}
   \phi(\vec{x})
=
   g
   \int \dr^{3}\vec{y} \,
   \langle \vec{x} |
   \biggl( \frac{1}{\vec{\partial}^{2}-m^{2}} \biggr)_{\vec{x}\vec{y}}
   | \vec{y} \rangle
   \Phi^{\dag}(\vec{y})\Phi(\vec{y}),
\label{eq:solution_phi}
\end{align}
by dropping the temporal dependence in $x^{\mu}=(x_{0},\vec{x})$ and $y^{\mu}=(y_{0},\vec{y})$.
The states $|x\rangle$ and $|y\rangle$ are also changed to $|\vec{x}\rangle$ and $|\vec{y}\rangle$, respectively.
Hereafter, we omit $x_{0}$ and $y_{0}$ 
unless required for specification.

From the Lagrangian~\eqref{eq:Lagrangian_simple}, we obtain
 the interaction Hamiltonian
$\displaystyle H_{\intn} = \int \dr^{4}x \, {\cal H}_{\intn}(x)$ with
$
   {\cal H}_{\intn}(x)
=
  g
  \phi(x) \Phi^{\dag}(x)\Phi(x)
$.
In the following discussion, we express this term by ${\cal H}_{\intn}(\vec{x}) = g \phi(\vec{x}) \Phi^{\dag}(\vec{x})\Phi(\vec{x})$ because the temporal dependence is dropped in the nonrelativistic approximation.
The expectation value of ${\cal H}_{\intn}(\vec{x})$ leads to the energy shift of the system:
\begin{align}
   \Delta E
&\equiv
   \langle 1, 2 |
   \int \dr^{3}\vec{x} \, {\cal H}_{\intn}(\vec{x})
   | 1, 2 \rangle,
\end{align}
with $|1,2\rangle=|1\rangle \otimes |2\rangle$ where $|1\rangle$ and $|2\rangle$ denote the heavy-particle states at the position 1 and 2, respectively, at the equal time.
By using Eq.~\eqref{eq:solution_phi},
we rewrite $\Delta E$ in the following form:
\begin{widetext}
\begin{align}
   \Delta E
&=
   g^{2}
   \int \dr^{3}\vec{x}
   \int \dr^{3}\vec{y} \,
   \langle 1,2 |
   \Phi^{\dag}(\vec{x})\Phi(\vec{x})
   \langle \vec{x} |
   \int \frac{\dr^{3}\vec{p}}{(2\pi)^{3}}
   | \vec{p} \rangle \langle \vec{p} |
   \frac{1}{\vec{\partial}^{2}-m^{2}}
   \int \frac{\dr^{3}\vec{q}}{(2\pi)^{3}}
   | \vec{q} \rangle \langle \vec{q}
   | \vec{y} \rangle
   \Phi^{\dag}(\vec{y})\Phi(\vec{y})
   | 1,2 \rangle
\nonumber \\[0.2em] 
&=
   \int \dr^{3}\vec{x}
   \int \dr^{3}\vec{y} \,
   \langle1| \Phi^{\dag}(\vec{x}) |0\rangle \langle0| \Phi(\vec{x}) |1\rangle
   \tilde{V}_{\phi}(\vec{x},\vec{y})
   \langle2| \Phi^{\dag}(\vec{y}) |0\rangle \langle0| \Phi(\vec{y}) |2\rangle.
\label{eq:Delta_E}
\end{align}
\end{widetext}
In the last equation, we have inserted the vacuum state denoted by $|0\rangle$ normalized by $\langle0|0\rangle=1$.
We have used $\langle \vec{x} | \vec{p} \rangle = e^{i\vec{p}\cdot\vec{x}}$ for the plane wave, and defined the potential by
\begin{align}
   \tilde{V}_{\phi}(\vec{x},\vec{y})
\equiv
   g^{2}
   \int \frac{\dr^{3}\vec{p}}{(2\pi)^{3}}
   \frac{-1}{\vec{p}^{2}+m^{2}}
   e^{-i\vec{p}\cdot(\vec{x}-\vec{y})},
\label{eq:tilde_V}
\end{align}
between $\vec{x}$ and $\vec{y}$.

Let us consider the scattering process $\vec{p}_{1}+\vec{p}_{2} \rightarrow \vec{p}_{1}'+\vec{p}_{2}'$ of two $\Phi$ particles, where the states $|1\rangle$ and $|2\rangle$ ($\langle1|$ and $\langle2|$) have the three-dimensional momenta $\vec{p}_{1}$ and $\vec{p}_{2}$ ($\vec{p}_{1}'$ and $\vec{p}_{2}'$), respectively.
Here we need to evaluate the wave functions, $\langle0| \Phi(\vec{x}) |1\rangle$, $\langle0| \Phi(\vec{y}) |2\rangle$, $\langle1| \Phi^{\dag}(\vec{x}) |0\rangle$, and $\langle2| \Phi^{\dag}(\vec{y}) |0\rangle$, in the plane waves with momentum $\vec{p}_{1}$, $\vec{p}_{2}$, $\vec{p}_{1}'$, and $\vec{p}_{2}'$.
For this purpose, we expand $\Phi(\vec{x})$ by
\begin{align}
   \Phi(\vec{x})
= \int \frac{\dr^{3}\vec{p}}{(2\pi)^{3}}
   \frac{1}{\sqrt{2E_{\vec{p}}}}
   \bigl( a_{\vec{p}}e^{i\vec{p}\cdot\vec{x}} + b_{\vec{p}}^{\dag}e^{-i\vec{p}\cdot\vec{x}} \bigr),
\label{eq:Phi_expansion}
\end{align}
according to the conventional forms,
where $E_{\vec{p}}=\sqrt{\vec{p}^{2}+M^{2}}$ is the energy of the heavy particle, and
$a_{\vec{p}}$ and $b_{\vec{p}}$ ($a_{\vec{p}}^{\dag}$ and $b_{\vec{p}}^{\dag}$) are the annihilation (creation) operators for the particle and antiparticle states with three-dimensional momentum $\vec{p}$.
The commutation relations for $a_{\vec{p}}$ and $a_{\vec{p}}^{\dag}$ ($b_{\vec{p}}$ and $b_{\vec{p}}^{\dag}$) are given by
$
   [a_{\vec{p}},a_{\vec{p}'}^{\dag}] = 
   [b_{\vec{p}},b_{\vec{p}'}^{\dag}] = (2\pi)^{3} \delta^{(3)}(\vec{p}-\vec{p}')
$.
In the followings, we consider only the particle state described by $a_{\vec{p}}$ and $a_{\vec{p}}^{\dag}$ by neglecting the antiparticle states.

We consider the state given by
$|\vec{p}\rangle=\sqrt{2E_{\vec{p}}}a_{\vec{p}}^{\dag}|0\rangle$.
The normalization of $|\vec{p}\rangle$ is given by
\begin{align}
   \langle\vec{p}|\vec{p}\rangle=2E_{\vec{p}}(2\pi)^{3}\delta^{3}(0) = 2E_{\vec{p}}V,
\label{eq:wavefunction_normalization}
\end{align}
which has the factor $2E_{\vec{p}}V$,
where $V$ is a volume of the whole space. 
This indicates that the number of the particle in the wave function is $2E_{\vec{p}}V$.
In Eq.~\eqref{eq:Delta_E}, we calculate $\langle0| \Phi(\vec{x}) |1\rangle$, $\langle0| \Phi(\vec{y}) |2\rangle$, $\langle1| \Phi^{\dag}(\vec{x}) |0\rangle$, and $\langle2| \Phi^{\dag}(\vec{y}) |0\rangle$.

We represent the states by $|\vec{p}_{1}\rangle$, $|\vec{p}_{2}\rangle$, $\langle\vec{p}_{1}'|$, and $\langle\vec{p}_{2}'|$, and consider  $\langle0| \Phi(\vec{x}) |\vec{p}_{1}\rangle$, $\langle0| \Phi(\vec{y}) |\vec{p}_{2}\rangle$, $\langle\vec{p}_{1}'| \Phi^{\dag}(\vec{x}) |0\rangle$, and $\langle\vec{p}_{2}'| \Phi^{\dag}(\vec{y}) |0\rangle$.
Using Eq.~\eqref{eq:Phi_expansion},
we obtain
\begin{align}
   \langle0|
 \Phi(\vec{x}) 
 |\vec{p}_{1}\rangle
&=
   e^{i\vec{p}_{1}\cdot\vec{x}},
   \\ 
   \langle0| \Phi(\vec{y}) |\vec{p}_{2}\rangle &= e^{i\vec{p}_{2}\cdot\vec{y}}, \\ 
   \langle\vec{p}_{1}'| \Phi^{\dag}(\vec{x}) |0\rangle &= e^{-i\vec{p}_{1}'\cdot\vec{x}}, \\ 
   \langle\vec{p}_{2}'| \Phi^{\dag}(\vec{y}) |0\rangle &= e^{-i\vec{p}_{2}'\cdot\vec{y}}.
\end{align}
Then, we find that $\Delta E$, which stems from $\Delta E$ in the relativistic version of the states, is expressed by
\begin{align}
   \Delta E
&=
   \int \dr^{3}\vec{x}
   \int \dr^{3}\vec{y} \,
   \tilde{V}_{\phi}(\vec{x},\vec{y})
   e^{i(\vec{p}_{1}-\vec{p}_{1}')\cdot\vec{x}}
   e^{i(\vec{p}_{2}-\vec{p}_{2}')\cdot\vec{y}}.
\end{align}
When we consider the limit of $\vec{p}_{1}, \vec{p}_{2}, \vec{p}_{1}', \vec{p}_{2}' \rightarrow 0$ 
in the static 
approximation,  
we express $\Delta E$ by
\begin{align}
   \Delta E
&\approx
   \int \dr^{3}\vec{x}
   \int \dr^{3}\vec{y} \,
   \tilde{V}_{\phi}(\vec{x},\vec{y}).
\end{align}
From Eq.~\eqref{eq:wavefunction_normalization}, we remember that the states $|\vec{p}_{1}\rangle$, $|\vec{p}_{2}\rangle$, $\langle\vec{p}_{1}'|$, and $\langle\vec{p}_{2}'|$ are normalized to have $2E_{\vec{p}_{1}}V$, $2E_{\vec{p}_{2}}V$, $2E_{\vec{p}_{1}'}V$, $2E_{\vec{p}_{2}'}V \approx 2MV$ particles in the nonrelativistic limit.
Then, we should regard the quantity $\Delta E/(2MV)^{2}$ as the potential energy for a pair of particles.
Thus, the energy per a pair of particles is given by
\begin{align}
   V_{\phi}(\vec{x},\vec{y})
&\equiv
   \frac{1}{(2M)^{2}}
   \tilde{V}_{\phi}(\vec{x},\vec{y}) ,
\end{align}
with $\tilde{V}_{\phi}(\vec{x},\vec{y})$ in Eq.~\eqref{eq:tilde_V}.
As a conclusion, $V_\phi$ is the potential between two $\Phi$'s used in the non-relativistic quantum mechanics.

\section{Derivation of OPEP for a $P^{(\ast)}$ meson and a nucleon} \label{sec:derivation_OPEP}

From Eqs.~\eqref{eq:LpiHH} and \eqref{eq:LpiNN_av}, we obtain the Lagrangian including $\pi$, $N$, and $H$ $(=P,P^\ast)$
\begin{align}
   {\cal L}_{\pi HN}
= \, &
   \frac{1}{2}
   \bigl( \partial_{\mu}\pi_{a} \partial^{\mu}\pi_{a} - m^{2}\pi_{a}^{2} \bigr)
   \nonumber \\ & 
+ \frac{ig_{\pi}}{f_{\pi}}
   \varepsilon_{\nu\rho\mu\sigma}v^{\nu}
   P_{\beta}^{\ast\rho\dag} \bigl(\vec{\tau} \!\cdot\! \partial^{\mu}\vec{\pi}\bigr)_{\beta\alpha} P_{\alpha}^{\ast\sigma}
   \nonumber \\ & 
 + i  \frac{ig_{\pi}}{f_{\pi}}
   P_{\beta\mu}^{\ast\dag} \bigl(\vec{\tau} \!\cdot\! \partial^{\mu}\vec{\pi}\bigr)_{\beta\alpha} P_{\alpha}
 \notag\\
 & + i  \frac{ig_{\pi}}{f_{\pi}}
   P_{\beta}^{\dag} \bigl(\vec{\tau} \!\cdot\! \partial^{\mu}\vec{\pi}\bigr)_{\beta\alpha} P_{\alpha\mu}^{\ast}
   \nonumber \\ & 
 +  \frac{g_{\pi NN}}{2m_{N}}
   \bar{\psi}_{\beta} \gamma_{\mu}\gamma_{5} (\vec{\tau}\!\cdot\!\partial^{\mu}\vec{\pi})_{\beta\alpha} \psi_{\alpha},
\label{eq:Lagrangian_pi_N_H}
\end{align}
where the kinetic terms of $H$ and $N$ are not shown.
The equation of motion for $\pi$ is 
\begin{widetext}
\begin{align}
   (\partial^{2} + m^{2}) \pi_{a}
=
 - \frac{ig_{\pi}}{f_{\pi}}
 \partial^{\mu}
 \Bigl(
 \varepsilon_{\nu\rho\mu\sigma}v^{\nu}
 P_{\beta}^{\ast\rho\dag}
 (\tau_{a})_{\beta\alpha}
 P_{\alpha}^{\ast\sigma}
 + i  P_{\beta\mu}^{\ast\dag}
 (\tau_{a})_{\beta\alpha}
 P_{\alpha}
 { + i} 
 P_{\beta}^{\dag}
 (\tau_{a})_{\beta\alpha}
 P_{\alpha\mu}^{\ast}
 \Bigr)
 { -} 
 \frac{g_{\pi NN}}{2m_{N}} 
 \partial_{\mu} \bigl( \bar{\psi}_{\beta}\gamma^{\mu}\gamma_{5} (\tau_{a})_{\beta\alpha} \psi_{\alpha} \bigr).
\end{align}
When we consider only the spatial dependence in the fields, we express the solution by
\begin{align}
   \pi_{a}(\vec{x})
= & 
 - \frac{ig_{\pi}}{f_{\pi}}
   \int \dr^{3} \vec{y} \,
   \langle \vec{x} |
   \frac{1}{-\vec{\partial}^{2}+m^{2}}
   | \vec{y} \rangle
   \partial_{y}^{j}
   \Bigl(
         \varepsilon_{\tau \chi j \omega}v^{\tau}
         P_{\delta}^{\ast\chi\dag}(\vec{y})
         (\tau_{a})_{\delta\gamma}
         P_{\gamma}^{\ast\omega}(\vec{y})
       { +i } 
         P_{\delta j}^{\ast\dag}(\vec{y})
         (\tau_{a})_{\delta\gamma}
         P_{\gamma}(\vec{y})
      { +i} 
         P_{\delta}^{\dag}(\vec{y})
         (\tau_{a})_{\delta\gamma}
         P_{\gamma j}^{\ast}(\vec{y})
   \Bigr)
   \nonumber \\ & 
{ -} 
 \frac{g_{\pi NN}}{2m_{N}}   
   \int \dr^{3} \vec{y} \,
   \langle \vec{x} |
   \frac{1}{-\vec{\partial}^{2}+m^{2}}
   | \vec{y} \rangle
   \partial_{yj} \bigl( \bar{\psi}_{\beta}(\vec{y})\gamma^{j}\gamma_{5} (\tau_{a})_{\beta\alpha} \psi_{\alpha}(\vec{y}) \bigr),
\end{align}
with $i,j=1,2,3$ for given $\psi$, $P$, and $P^{\ast}$.
Then, the interaction energy between $P^{(\ast)}$ and $N$ is given by
\begin{align}
   \Delta E^{HN}
\equiv \, &
   \langle1,2|
   \int \dr^{3} \vec{x} \,
   {\cal H}_{\intn}^{\pi H N}(\vec{x})
   |1,2\rangle
\nonumber \\[0.2em] 
= \, &
   \int \dr^{3} \vec{x}
   \int \dr^{3} \vec{y} \,
   \langle1|
   \Bigl(
       - \varepsilon_{ikl}
         P_{\beta}^{\ast k \dag}(\vec{x})
         (\tau_{a})_{\beta\alpha}
         P_{\alpha}^{\ast l}(\vec{x})
       { +i} 
         P_{\beta i}^{\ast\dag}(\vec{x})
         (\tau_{a})_{\beta\alpha}
         P_{\alpha}(\vec{x})
      { +i} 
         P_{\beta}^{\dag}(\vec{x})
         (\tau_{a})_{\beta\alpha}
         P_{\alpha i}^{\ast}(\vec{x})
   \Bigr)
   |1\rangle
   \nonumber \\ & \times 
   \tilde{V}_{\pi \, ij}^{HN}(\vec{x},\vec{y})
   \langle2|
   \bar{\psi}_{\beta'}(\vec{y})\gamma^{j}\gamma_{5} (\tau_{a})_{\beta'\alpha'} \psi_{\alpha'}(\vec{y})
   |2\rangle,
\end{align}
where ${\cal H}_{\intn}^{\pi H N}$ represents the interaction Hamiltonian stemming from Eq.~\eqref{eq:Lagrangian_pi_N_H}, and $|1\rangle$ and $|2\rangle$ represent a $P^{(\ast)}$ meson and a nucleon, respectively.
For brevity we have defined
$\tilde{V}_{\pi \, ij}^{HN}(\vec{x},\vec{y})$
by
\begin{align}
   \tilde{V}_{\pi \, ij}^{HN}(\vec{x},\vec{y})
&\equiv
 { -} 
   \frac{g_{\pi NN}}{2m_{N}}
   \frac{ig_{\pi}}{f_{\pi}}
   \int \frac{\dr^{3}\vec{p}}{(2\pi)^{3}}
   \frac{p_{i}p_{j}}{\vec{p}^{2}+m^{2}}
   e^{-i\vec{p}\cdot(\vec{x}-\vec{y})}.
\end{align}
\end{widetext}
We consider the matrix element by using the basis states
$|1\rangle=|P_{\alpha_{1}}^{\ast}(\vec{p}_{1},\lambda_{1})\rangle$ or $|P_{\alpha_{1}}(\vec{p}_{1})\rangle$ and $\langle1|=\langle P_{\beta_{1}}^{\ast}(\vec{p}_{1}',\lambda_{1}')|$ or $\langle P_{\beta_{1}}(\vec{p}_{1}')|$.
Here $\vec{p}_{1}$ ($\vec{p}_{1}'$) is the three-dimensional momentum of the $P^{(\ast)}$ meson and $\lambda_{1}$ ($\lambda_{1}'$) is the helicity of the $P^{\ast}$ meson ($\lambda_{1},\lambda_{1}'=0,\pm$).
$\alpha_{1},\beta_{1}=\pm1/2$ are the isospin components.
Adopting the following channels,
\begin{align}
   \bigl\{ \langle1|, |1\rangle \bigr\}
=&
   \bigl\{ \langle P_{\beta_{1}}^{\ast}(\vec{p}_{1}',\lambda_{1}')|, | P_{\alpha_{1}}^{\ast}(\vec{p}_{1},\lambda_{1})\rangle \bigr\},
   \nonumber \\ & 
   \bigl\{ \langle P_{\beta_{1}}^{\ast}(\vec{p}_{1}',\lambda_{1}')|, | P_{\alpha_{1}}(\vec{p}_{1})\rangle \bigr\},
   \nonumber \\ & 
   \bigl\{ \langle P_{\beta_{1}}(\vec{p}_{1}')|, | P_{\alpha_{1}}^{\ast}(\vec{p}_{1},\lambda_{1})\rangle \bigr\},
\end{align}
and
\begin{align}
   \bigl\{ \langle2|, |2\rangle \bigr\}
=&
   \bigl\{ \langle N_{\beta_{2}}(\vec{p}_{2}',s_{2}') |, | N_{\alpha_{2}}(\vec{p}_{2},s_{2}) \rangle \bigr\},
\end{align}
we obtain the potential energy in each channel:
\begin{widetext}
\begin{align}
   \Delta E_{P^{\ast}N\text{-}P^{\ast}N}
= &
   \int \dr^{3} \vec{x}
   \int \dr^{3} \vec{y} \,
   \langle\bar{D}_{\beta_{1}}^{\ast}(\vec{p}_{1}',\lambda_{1}')|
   \Bigl(
       - \varepsilon_{ikl}
         P_{\beta}^{\ast k \dag}(\vec{x})
         (\tau_{a})_{\beta\alpha}
         P_{\alpha}^{\ast l}(\vec{x})
   \Bigr)
   |\bar{D}_{\alpha_{1}}^{\ast}(\vec{p}_{1},\lambda_{1})\rangle
   \tilde{V}_{\pi ij}^{HN}(\vec{x},\vec{y})
   \nonumber \\ & \times 
   \langle N_{\beta_{2}}(\vec{p}_{2}',s_{2}') |
   \bar{\psi}_{\beta'}(\vec{y})\gamma^{j}\gamma_{5} (\tau_{a})_{\beta'\alpha'} \psi_{\alpha'}(\vec{y})
   | N_{\alpha_{2}}(\vec{p}_{2},s_{2}) \rangle,
 \label{eq:energies_P*_P*_N}
\\[0.2em] 
   \Delta E_{P^{\ast}N\text{-}PN}
= &
   \int \dr^{3} \vec{x}
   \int \dr^{3} \vec{y} \,
   \langle\bar{D}_{\beta_{1}}^{\ast}(\vec{p}_{1}',\lambda_{1}')|
   \Bigl(
       i  P_{\beta i}^{\ast\dag}(\vec{x})
         (\tau_{a})_{\beta\alpha}
         P_{\alpha}(\vec{x})
   \Bigr)
   |\bar{D}_{\alpha_{1}}(\vec{p}_{1})\rangle
   \tilde{V}_{\pi ij}^{HN}(\vec{x},\vec{y})
   \nonumber \\ & \times 
   \langle N_{\beta_{2}}(\vec{p}_{2}',s_{2}') |
   \bar{\psi}_{\beta'}(\vec{y})\gamma^{j}\gamma_{5} (\tau_{a})_{\beta'\alpha'} \psi_{\alpha'}(\vec{y})
   | N_{\alpha_{2}}(\vec{p}_{2},s_{2}) \rangle,
\\[0.2em] 
   \Delta E_{PN\text{-}P^{\ast}N}
= &
   \int \dr^{3} \vec{x}
   \int \dr^{3} \vec{y} \,
   \langle\bar{D}_{\beta_{1}}(\vec{p}_{1}')|
   \Bigl(
        i P_{\beta}^{\dag}(\vec{x})
         (\tau_{a})_{\beta\alpha}
         P_{\alpha i}^{\ast}(\vec{x})
   \Bigr)
   |\bar{D}_{\alpha_{1}}^{\ast}(\vec{p}_{1},\lambda_{1})\rangle
   \tilde{V}_{\pi ij}^{HN}(\vec{x},\vec{y})
   \nonumber \\ & \times 
   \langle N_{\beta_{2}}(\vec{p}_{2}',s_{2}') |
   \bar{\psi}_{\beta'}(\vec{y})\gamma^{j}\gamma_{5} (\tau_{a})_{\beta'\alpha'} \psi_{\alpha'}(\vec{y})
   | N_{\alpha_{2}}(\vec{p}_{2},s_{2}) \rangle .
\label{eq:energies_P_P*_N}
\end{align}
\end{widetext}
Here $\vec{p}_{2}$ and $\vec{p}_{2}'$ are the three-dimensional momenta of the nucleon, $s_{2},s_{2}'=\pm1/2$ are 
the spin components, and $\alpha_{2},\beta_{2}=\pm1/2$ are the isospin components.

In order to calculate the matrix elements,
we expand $P_{\alpha}^{\ast i}(\vec{x})$ and $P_{\alpha}(\vec{x})$ by plane waves.
This is obtained by considering 
multiplying
the mass scale $M$ to Eq.~\eqref{eq:Phi_expansion} and taking the large $M$ limit.
The results are
\begin{align}
   P^{\ast i}_{\alpha}(\vec{x})
&=
   \int \frac{\dr^{3}\vec{p}}{(2\pi)^{3}}
   \frac{1}{\sqrt{2}}
   \bigl( a_{\vec{p}\alpha}^{i}e^{i\vec{p}\cdot\vec{x}} + b_{\vec{p}\alpha}^{i\dag}e^{-i\vec{p}\cdot\vec{x}} \bigr),
\label{eq:P*_expansions} \\ 
   P_{\alpha}(\vec{x})
&=
   \int \frac{\dr^{3}\vec{p}}{(2\pi)^{3}}
   \frac{1}{\sqrt{2}}
   \bigl( a_{\vec{p}\alpha}e^{i\vec{p}\cdot\vec{x}} + b_{\vec{p}\alpha}^{\dag}e^{-i\vec{p}\cdot\vec{x}} \bigr),
\label{eq:P_expansions}
\end{align}
where the factor $1/\sqrt{2}$ stems from $1/\sqrt{2E_{\vec{p}}}$ in the conventional representation multiplied by the factor $\sqrt{M}$ and taking the large $M$ limit.
Here $a_{\vec{p}\alpha}^{i}$ and $b_{\vec{p}\alpha}^{i}$ ($a_{\vec{p}\alpha}$ and $b_{\vec{p}\alpha}$) satisfy the commutation relations, 
$
   [a_{\vec{p}}^{i},a_{\vec{p}'}^{j\dag}] =
   [b_{\vec{p}}^{i},b_{\vec{p}'}^{j\dag}] = (2\pi)^{3} \delta^{ij} \delta^{(3)}(\vec{p}-\vec{p}')
$ and
$
   [a_{\vec{p}},a_{\vec{p}'}^{\dag}] =
   [b_{\vec{p}},b_{\vec{p}'}^{\dag}] = (2\pi)^{3} \delta^{(3)}(\vec{p}-\vec{p}')
$.
Let us consider the large $M$ limit and leave only the leading terms of $M$.
Because the particle states
$|P_{\alpha}^{\ast}(\vec{p},\lambda)\rangle$ and $|P_{\alpha}(\vec{p})\rangle$
are defined by
\begin{align}
   |P_{\alpha}^{\ast}(\vec{p},\lambda)\rangle
&\equiv
   \sqrt{2}
   \epsilon_{i}^{(\lambda)} (a_{\vec{p}\alpha}^{i})^{\dag}
   |0\rangle,
\\ 
   |P_{\alpha}(\vec{p})\rangle
&\equiv
   \sqrt{2}
   (a_{\vec{p}\alpha}^{i})^{\dag}
   |0\rangle,
\label{eq:P_states_HQS}
\end{align}
which indicate that the states
$|P_{\alpha}^{\ast}(\vec{p},\lambda)\rangle$ and $|P_{\alpha}(\vec{p})\rangle$
include $2V$ particles in the volume $V$.
The polarization vectors for the $P^{\ast}$ meson are given by Eq.~\eqref{eq:polarization_vector}.
We also use $T_{i}$ ($i=1,2,3$) in Eq.~\eqref{eq:T_matrices}.
As for the nucleon part, we consider the expansion for $\psi(\vec{x})$ given by
\begin{align}
   \psi(\vec{x})
&=
   \int \frac{\dr^{3}\vec{p}}{(2\pi)^{3}}
   \frac{1}{\sqrt{2E_{\vec{p}}}}
   \nonumber \\ & \times 
   \sum_{s=\pm1/2}
   \sum_{\alpha=\pm1/2}
   \Bigl(
         a_{\vec{p}s\alpha} u_{s}(\vec{p}) e^{i\vec{p}\cdot\vec{x}}
      + b_{\vec{p}s\alpha} v_{s}(\vec{p}) e^{-i\vec{p}\cdot\vec{x}}
   \Bigr),
\end{align}
with the commutation relations
\begin{align}
   [a_{\vec{p}r\beta},a_{\vec{q}s\alpha}^{\dag}]
=
   [b_{\vec{p}r\beta},b_{\vec{q}s\alpha}^{\dag}]
=
   (2\pi)^{3} \delta^{(3)}(\vec{p}-\vec{q}) \delta_{rs} \delta_{\beta\alpha},
\end{align}
 for spin $s,r$ and isospin $\alpha,\beta$.
The normalizations for $u$ and $v$ by $u_{r}(\vec{p})^{\dag}u_{s}(\vec{p})=v_{r}(\vec{p})^{\dag}v_{s}(\vec{p})=2E_{\vec{p}}\xi_{r}^{\dag}\xi_{s}$ and $\bar{u}_{r}(\vec{p})u_{s}(\vec{p})=-\bar{v}_{r}(\vec{p})v_{s}(\vec{p})=2m_{N}\xi_{r}^{\dag}\xi_{s}$ with $E_{\vec{p}}=\sqrt{\vec{p}^{2}+m_{N}^{2}}$ for the nucleon mass $m_{N}$.
The concrete forms of $u_{s}(\vec{p})$ and $v_{s}(\vec{p})$ are given by
\begin{align}
   u_{s}(\vec{p})
&=
   \sqrt{E_{\vec{p}}+m_{N}}
   \left(
   \begin{array}{c}
     \xi_{s} \\
     \dfrac{\vec{p}\!\cdot\!\vec{\sigma}}{E_{\vec{p}}+m_{N}} \xi_{s}
   \end{array}
   \right),
   \\ 
   v_{s}(\vec{p})
&=
   \sqrt{E_{\vec{p}}+m_{N}}
   \left(
   \begin{array}{c}
     \dfrac{\vec{p}\!\cdot\!\vec{\sigma}}{E_{\vec{p}}+m_{N}} \zeta_{s} \\
     \zeta_{s}
   \end{array}
   \right),
\end{align}
for the standard representation of the Dirac matrices,
where $u_{s}(\vec{p})^{\dag}u_{s}(\vec{p})=v_{s}(\vec{p})^{\dag}v_{s}(\vec{p})=2E_{\vec{p}}$ holds for the normalization $|\xi_{s}|^{2}=|\zeta_{s}|^{2}=1$.
We consider the scattering process for the nucleons, $(\vec{p}_{1},s_{1},\alpha_{1})+(\vec{p}_{2},s_{2},\alpha_{2}) \rightarrow (\vec{p}_{1}',s_{1}',\alpha_{1}')+(\vec{p}_{2}',s_{2}',\alpha_{2}')$, with $\vec{p}_{i}$ ($\vec{p}_{i}'$) for the initial (final) momentum, $s_{i}$ and $\alpha_{i}$ ($s_{i}'$ and $\alpha_{i}'$) for the initial (final) spin and isospin for the nucleon $i=1,2$.
The wave functions are denoted by $|1\rangle=|\vec{p}_{1}s_{1}\alpha_{1}\rangle$, $|2\rangle=|\vec{p}_{2}s_{2}\alpha_{2}\rangle$, $\langle1|=\langle\vec{p}_{1}'s_{1}'\alpha_{1}'|$, and $\langle2|=\langle\vec{p}_{2}'s_{2}'\alpha_{2}'|$.
We define the plane-wave state by
\begin{align}
   |N_{s\alpha}(\vec{p})\rangle \equiv \sqrt{2E_{\vec{p}}} \, a_{\vec{p}s\alpha}^{\dag} |0\rangle,
\end{align}
for the vacuum state $|0\rangle$ with the normalization $\langle0|0\rangle=1$.
The normalization for $|\vec{p}s\alpha\rangle$ is given by
$
   \langle\vec{p}'s'\alpha'|\vec{p}s\alpha\rangle
= 2E_{\vec{p}} (2\pi)^{3}\delta^{(3)}(\vec{p}'-\vec{p}) \delta_{s's} \delta_{\alpha'\alpha}
$.

From 
Eqs.~\eqref{eq:energies_P*_P*_N}-\eqref{eq:energies_P_P*_N},
we obtain the potentials 
\begin{widetext}
\begin{align}
   V_{\pi}^{P^{\ast}N\text{-}P^{\ast}N}(\vec{x},\vec{y})
&=
 { -}
   { \frac{1}{2}}
   \frac{g_{\pi NN}}{2m_{N}}
   \frac{g_{\pi}}{f_{\pi}}
   \int \frac{\dr^{3}\vec{p}}{(2\pi)^{3}}
   \frac{p_{i}p_{j}}{\vec{p}^{2}+m^{2}}
   e^{-i\vec{p}\cdot(\vec{x}-\vec{y})}
   (T_{i})_{\lambda_{1}'\lambda_{1}}
   (\sigma_{j})_{s_{2}'s_{2}}
   (\tau_{a})_{\beta_{1}\alpha_{1}}
   (\tau_{a})_{\beta_{2}\alpha_{2}},
\label{eq:barDN_V_PS_potentials_3a} \\[0.5em] 
   V_{\pi}^{P^{\ast}N\text{-}PN}(\vec{x},\vec{y})
&=
 \frac{1}{2}
   \frac{g_{\pi NN}}{2m_{N}}
   \frac{g_{\pi}}{f_{\pi}}
   \int \frac{\dr^{3}\vec{p}}{(2\pi)^{3}}
   \frac{p_{i}p_{j}}{\vec{p}^{2}+m^{2}}
   e^{-i\vec{p}\cdot(\vec{x}-\vec{y})}
   \epsilon_{i}^{(\lambda_{1}')*}
   (\sigma_{j})_{s_{2}'s_{2}}
   (\tau_{a})_{\beta_{1}\alpha_{1}}
   (\tau_{a})_{\beta_{2}\alpha_{2}},
\label{eq:barDN_V_PS_potentials_3b} \\[0.5em] 
   V_{\pi}^{PN\text{-}P^{\ast}N}(\vec{x},\vec{y})
&=
 \frac{1}{2}
   \frac{g_{\pi NN}}{2m_{N}}
   \frac{g_{\pi}}{f_{\pi}}
   \int \frac{\dr^{3}\vec{p}}{(2\pi)^{3}}
   \frac{p_{i}p_{j}}{\vec{p}^{2}+m^{2}}
   e^{-i\vec{p}\cdot(\vec{x}-\vec{y})}
   \epsilon_{i}^{(\lambda_{1})}
   (\sigma_{j})_{s_{2}'s_{2}}
   (\tau_{a})_{\beta_{1}\alpha_{1}}
   (\tau_{a})_{\beta_{2}\alpha_{2}},
\label{eq:barDN_V_PS_potentials_3c}
\end{align}
\end{widetext}
 to be transformed to Eqs.~\eqref{eq:barDN_V_PS_potentials_3A}, \eqref{eq:barDN_V_PS_potentials_3B}, and \eqref{eq:barDN_V_PS_potentials_3C} in the end.
Notice 
that 
the factor $1/2$ in the coefficients 
have been missed in the previous studies by the authors~\cite{Yasui:2009bz,Yamaguchi:2011xb,Yamaguchi:2011qw}. 
The calculation of the momentum integrations is easily performed by introducing the form factor~\eqref{eq:dipole_form_factor} in the integrands.
In the calculations, it is useful to adopt the formula
 of the plane-wave expansion
\begin{align}
   e^{-i\vec{p}\cdot\vec{r}}
=
   4\pi \sum_{l,l_{z}} (-i)^{l} j_{l}(pr) Y_{ll_{z}}^{\ast}(\hat{\vec{p}}) Y_{ll_{z}}(\hat{\vec{r}}),
\end{align}
with $l=0,1,2,\dots$ and $l_{z}=-l,-l+1,\dots,l-1,l$.
Here $j_{l}(x)$ is the spherical Bessel function and $Y_{ll_{z}}(\hat{\vec{x}})$ is the spherical harmonic function.
As a result, we obtain the explicit forms of the central potential
$C(r;m)$ and the tensor potential
$T(r;m)$ in Eqs.~\eqref{eq:C_def} and \eqref{eq:T_def}, respectively.
In the calculation of the tensor potential, we have used the relationship
\begin{align}
   a_{i}b_{j}S_{ij}(\hat{\vec{p}})
&=
   \sqrt{\frac{24\pi}{5}}
   \sum_{\mu=-2}^{2}
   (-1)^{\mu}
   (\vec{a}\times\vec{b})^{(2)}_{\mu} Y_{2\mu}(\hat{\vec{p}}),
\end{align}
where $(\vec{a}\times\vec{b})^{(2)}_{\mu}$ is the rank-2 tensor composed of $\vec{a}=(a_{1},a_{2},a_{3})$ and $\vec{b}=(b_{1},b_{2},b_{3})$.

\bibliography{references}

\end{document}